\documentclass[journal]{IEEEtran}
\usepackage{subfigure}
\usepackage{booktabs}
\usepackage{marvosym}
\usepackage{amsmath,epsfig}
\usepackage{multirow}
\usepackage{url}
\usepackage{color}
\usepackage{amsfonts,amssymb}
\usepackage{amsmath,amsfonts}
\usepackage{arydshln}
\usepackage{multirow}
\ifCLASSINFOpdf
\else
\fi
\begin{document}
\title{Advancing Zero-Shot Digital Human Quality Assessment through Text-Prompted Evaluation}

\author{Zicheng Zhang, Wei Sun, Yingjie Zhou, Haoning Wu, Chunyi Li, Xiongkuo Min, \\   Xiaohong Liu, Guangtao Zhai, \emph{Senior Member, IEEE}, and Weisi Lin, \emph{Fellow, IEEE} 

\IEEEcompsocitemizethanks{\IEEEcompsocthanksitem Zicheng Zhang, Wei Sun, Yingjie Zhou, Chunyi Li, Xiongkuo Min, and Guangtao Zhai are with the Institute of Image Communication and Network Engineering, Shanghai Jiao Tong University, 200240 Shanghai, China. E-mail:\{zzc1998, sunguwei, zhouyingjie, lcysyzxdxc,
minxiongkuo, zhaiguangtao\}
@sjtu.edu.cn.\protect }

\IEEEcompsocitemizethanks{\IEEEcompsocthanksitem Haoning Wu is with S-Lab, Nanyang Technological University, Singapore. E-mail: haoning001@e.ntu.edu.sg.\protect}

\IEEEcompsocitemizethanks{\IEEEcompsocthanksitem Xiaohong Liu is with John Hopcroft Center, Shanghai Jiao Tong University, Shanghai 200240, China. E-mail: xiaohongliu@sjtu.edu.cn.\protect }

\IEEEcompsocitemizethanks{\IEEEcompsocthanksitem Weisi Lin is with School of Computer Science and Engineering, Nanyang Technological University, Singapore.  E-mail: wslin@ntu.edu.sg.\protect}
\thanks{(Corresponding author: Guangtao Zhai.) }
}

\maketitle

\begin{abstract}
Digital humans have witnessed extensive applications in various domains, necessitating related quality assessment studies. However, there is a lack of comprehensive digital human quality assessment (DHQA) databases. To address this gap, we propose SJTU-H3D, a subjective quality assessment database specifically designed for full-body digital humans. It comprises 40 high-quality reference digital humans and 1,120 labeled distorted counterparts generated with seven types of distortions. The SJTU-H3D database can serve as a benchmark for DHQA research, allowing evaluation and refinement of processing algorithms. Further, we propose a zero-shot DHQA approach that focuses on no-reference (NR) scenarios to ensure generalization capabilities while mitigating database bias. Our method leverages semantic and distortion features extracted from projections, as well as geometry features derived from the mesh structure of digital humans. Specifically, we employ the Contrastive Language-Image Pre-training (CLIP) model to measure semantic affinity and incorporate the Naturalness Image Quality Evaluator (NIQE) model to capture low-level distortion information. Additionally, we utilize dihedral angles as geometry descriptors to extract mesh features. By aggregating these measures, we introduce the Digital Human Quality Index (DHQI), which demonstrates significant improvements in zero-shot performance. The DHQI can also serve as a robust baseline for DHQA tasks, facilitating advancements in the field. The database and the code are available at https://github.com/zzc-1998/SJTU-H3D.
\end{abstract}

\begin{IEEEkeywords}
Digital humans, quality assessment, database, zero-shot, no-reference 
\end{IEEEkeywords}

% \fi
\IEEEpeerreviewmaketitle

\section{Introduction}
\IEEEPARstart{D}igital humans are computer-based simulations and models of human beings, extensively utilized in various applications such as gaming, the automotive industry, and the metaverse. The current research endeavors primarily focus on the generation, representation, rendering, and animation of digital humans \cite{zhu2019applications}. However, with the rapid advancement of virtual reality (VR) and augmented reality (AR) technologies, there is an increasing demand from users for higher visual quality of digital humans. Consequently, it has become imperative to conduct quality assessment studies on digital humans.
Regrettably, acquisition of digital human models is a laborious and costly process compared with 2D media such as images and videos, requiring specialized three-dimensional (3D) scanning devices and professional post-production, which makes it quite difficult to carry out digital human quality assessment (DHQA) databases. Therefore, few works about subjective DHQA have been carried out in the literature. Then the absence of large-scale subjective experiments for assessing the visual quality of digital humans further hinders progress in this domain. 

\begin{figure}
    \centering
    \includegraphics[width=\linewidth]{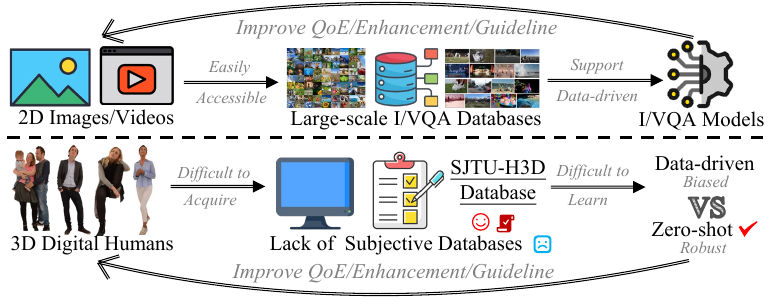}
    \caption{Motovation of our works. Unlike 2D images/videos, collection of 3D digital humans is more difficult and expensive. Therefore there is a lack of subjective databases for 3D digital humans currently. To tackle this issue, we propose the first perceptual quality assessment database for full-body digital humans (SJTU-H3D). Furthermore, in contrast to the numerous large-scale image/video quality assessment (I/VQA) databases that facilitate data-driven methodologies, supervised methods can be easily bothered by the bias of the limited number of DHQA databases, affecting generalization ability. Thus we propose a zero-shot no-reference quality assessment method to address this concern.}
    \label{fig:spotlight}
    \vspace{-0.3cm}
\end{figure}

Therefore, in this paper, we propose a comprehensive subjective quality assessment database (\textbf{SJTU-H3D}) targeted at digital humans, aiming to address this research gap and contribute to the advancement of DHQA. The SJTU-H3D database introduced in this study comprises 40 high-quality reference digital humans, represented by textured meshes in full-body format, and the database includes 1,120 distorted digital humans that have been generated using seven different types of distortions. The perceptual mean opinion scores (MOSs) of these distorted digital humans are  collected through a meticulously controlled subjective experiment.
Notably, the SJTU-H3D database is the first large-scale database specifically designed for digital human quality assessment (DHQA) that focuses on full-body representations. The primary objective of this database is to advance the research and development of DHQA within the scientific community. Furthermore, it serves as an ideal platform for evaluating and refining various processing algorithms, including but not being limited to denoising and compression techniques.
By providing a comprehensive database consisting of high-quality reference models and distorted counterparts, the proposed SJTU-H3D database offers researchers and practitioners an opportunity to explore and enhance their DHQA methodologies. The availability of such a resource is expected to significantly contribute to the growth and advancement of the DHQA research community.

% In past years, data-driven I/VQA approaches have been widely studied and achieved outstanding performance in various application scenarios. Benefited from the large-scale I/VQA databases (the SPAQ \cite{fang2020perceptual} database contains 11,125 labeled images and the LSVQ \cite{ying2021patch} databases contains even up to 38,811 annotated videos), the generalization ability and robustness of the data-driven approaches can be well confirmed. Unfortunately, in the field of DHQA research, only one perceptual quality assessment database (DHHQA \cite{zhang2023perceptual}) except the proposed SJTU-H3D database is carried out and focuses on the digital human heads rather than the full body. This makes it difficult to develop data-driven DHQA  methods and ensure such methods' generalization ability. 

% Therefore, the challenge motivates us to design a zero-shot DHQA method that requires no training on the labeled DHQA databases. To fit most practical applications, we only focus on the no-reference (NR) methods since pristine references are not always available. 

During recent years, data-driven image and video quality assessment (I/VQA) approaches \cite{zhang2023aigc,li2023agiqa,dong2023light, gao2023vdpve} have garnered significant attention and have demonstrated remarkable performance in various application domains. The success of these approaches can be partly attributed to the availability of large-scale I/VQA databases such as the SPAQ database (containing 11,125 labeled images) \cite{fang2020perceptual} and the LSVQ database (comprising up to 38,811 annotated videos) \cite{ying2021patch}. These databases have also contributed to ensuring the generalization capability and robustness of data-driven methods.
However, in the realm of DHQA research, the availability of suitable perceptual quality assessment databases is limited. With the exception of the proposed SJTU-H3D database, only one perceptual quality assessment database, DHHQA \cite{zhang2023perceptual}, focusing solely on digital human heads rather than full-body representations, exists. 
This scarcity of databases makes it challenging to develop data-driven DHQA methods and ensure their generalization ability in practical scenarios.

Hence, this challenge serves as a motivation for us to devise a zero-shot DHQA method that does not necessitate training on labeled DHQA databases. To cater to most practical applications where pristine references may not be readily available, our focus is only on no-reference (NR) methods. 
To extract both semantic and distortion features for evaluating the visual quality of digital humans, we employ projection rendering techniques. From a semantic perspective, we utilize the Contrastive Language-Image Pre-training (CLIP) model \cite{radford2021learning} to measure the correlation between the input projections and quality-related texts. Our hypothesis is that high-quality digital human projections should exhibit a strong correlation with positive quality-related texts and a weak correlation with negative ones. To determine the quality levels of the input projections, we design several positive-negative text pairs. The semantic affinity quality measure is then derived by computing the difference in affinity between positive and negative texts.
However, CLIP operates on low-resolution images, which limits its ability to capture low-level distortion information. To address this limitation, we incorporate the completely blind Naturalness Image Quality Evaluator (NIQE) \cite{mittal2012making} to extract low-level quality representations from the raw resolution. 
To further enhance the accuracy of quality prediction, we also extract features from the mesh modality. For robustness and effectiveness, we choose the dihedral angle as the geometry descriptor, as it has been widely recognized for effectively capturing geometric features relevant to visual quality \cite{alexiou2018point,nr-svr,dame,zhang2021mesh} and its values are confined within the range of [0, $\pi$]. By analyzing the changing tendency of dihedral angles corresponding to geometry compression and simplification levels, we average-pool the dihedral angles to derive the geometry loss quality measure.
Finally, all three quality measures (semantic affinity quality measure, spatial naturalness quality measure, and geometry loss quality measure) are aggregated using a sum function to form the proposed Digital Human Quality Index (DHQI). Experimental results demonstrate that DHQI significantly improves zero-shot performance and even achieves competitiveness with supervised methods. In summary, our contributions are as follows:
\begin{itemize}
    \item We propose the first large-scale full-body DHQA database, SJTU-H3D, which consists of 40 high-quality digital humans represented by textured meshes and 1,120 distorted digital humans generated by 7 types of distortions.
    We carry out a well-controlled subjective experiment. 40 human subjects are invited and a total of 44,800 ratings are collected to gather the mean opinion scores (MOSs) for 1,120 distorted digital humans.
    \item  We propose a novel text-prompted zero-shot digital human quality index. Extensive experiments are conducted, which include the general performance comparison, detailed distortion-specific performance comparison, statistical test, and ablation studies.
\end{itemize}

\section{Related Works}
In this section, we give a brief introduction to the development of 3D model quality assessment (3DQA) and no-reference image quality assessment (NR-IQA) methods.

\begin{table*}[!htp]
\centering
\caption{The comparison of current 3DQA databases and our database.}
\begin{tabular}{lcccc}
\toprule
Database        & Source   & Rated Models  &Format   & Content            \\
\midrule
SJTU-PCQA (TMM, 2020) \cite{yang2020predicting}    &10 & 378  & Colored Point Cloud  & Humans, Statues          \\
WPC (TVCG, 2022) \cite{liu2022perceptual}   &20 &740  &  Colored Point Cloud   & Fruit, Vegetables, Tools           \\
LSPCQA (TOMM, 2022) \cite{liu2022point} &104 &1,240    &Colored Point Cloud   & Animals, Humans, Vehicles, Daily objects  \\
CMDM (TVCG, 2021) \cite{y2021visual}  &5 & 80  & Colored Mesh   &   Humans, Animals, Statues  \\
TMQA (TOG, 2022) \cite{nehme2022textured} &55 &3,000 & Textured Mesh & Statues, Animals, Daily objects\\
DHHQA (ICASSP, 2023) \cite{zhang2023perceptual}           & 55  & 1,540 & Textured Mesh  & Scanned Real Human Heads\\
\textbf{SJTU-H3D (Ours, 2023)} & 40 & 1,120 & Textured Mesh & Full-body Digital Humans\\
\bottomrule
\end{tabular}
% \vspace{-0.3cm}
\vspace{-0.4cm}
\label{tab:comparison}
\end{table*}
\subsection{3DQA Development}
% 3D quality model assessment (3DQA) mainly focuses on point cloud quality assessment (PCQA) and mesh quality assessment (MQA). 

\subsubsection{3DQA Databases}
Early subjective 3D quality assessment (3DQA) databases primarily employ colorless point clouds and are relatively 
small in scale \cite{alexiou2017performance,alexiou2017towards,alexiou2018impact}. However, recent efforts have been directed towards addressing the challenge of assessing visual quality in colored 3D models, resulting in the development of substantial 3DQA databases \cite{alexiou2017performance,alexiou2017towards,alexiou2018impact,yang2020predicting,liu2022perceptual,liu2022point,y2021visual,nehme2022textured}. A detailed comparison between these databases and the proposed database is presented in Table \ref{tab:comparison}. From the table, it is evident that the recent 3DQA databases, with the exception of DHHQA, encompass general 3D objects and do not specifically focus on 3D digital humans. Although the DHHQA database comprises real human heads, it neglects the consideration of the body part. This highlights the significance of the proposed SJTU-H3D database. 

\subsubsection{3DQA Methods}

In the field of 3D quality assessment (3DQA), metrics can be broadly categorized into model-based and projection-based methods. Model-based methods \cite{mekuria2016evaluation,tian2017geometric,alexiou2018point,torlig2018novel,yang2020graphsim,meynet2020pcqm,alexiou2020pointssim,pcqa-large-scale,zhang2022no,zhou2022blind} involve extracting features directly from the 3D model, which offers the advantage of being viewpoint-invariant and relatively straightforward. However, due to the inherent complexity of 3D models, these methods can be computationally expensive and time-consuming.
On the other hand, projection-based methods \cite{yang2020predicting,liu2021pqa,zhang2021mesh,fan2022no,zhang2022treating} infer the visual quality of a 3D model based on its corresponding projections. These methods leverage mature and successful 2D media analysis tools, which often lead to excellent performance. However, projection-based methods are highly dependent on the selection of viewpoints and can be susceptible to instability when subjected to various rendering setups.
More recently, some 3DQA methods \cite{zhang2023mm} have emerged that aim to combine features from multiple modalities. By leveraging the advantages of both model-based and projection-based modalities, these methods attempt to enhance overall performance. This integration of features from different modalities allows for a more comprehensive assessment of 3D model quality.

\subsection{NR-IQA Methods}
\subsubsection{Supervised NR-IQA}
In general, NR-IQA methods aim to evaluate image quality without reference information, and they can be classified into handcrafted-based methods, which extract features using manual techniques, and deep learning-based methods, which employ deep neural networks for feature extraction, both of which have demonstrated effectiveness in common IQA tasks.
One representative handcrafted-based method is BRISQUE \cite{mittal2012brisque}, which utilizes natural scene statistics (NSS) in the spatial domain to analyze image quality. CPBD \cite{narvekar2011no} estimates blur levels by computing the cumulative probability of blur detection. BMPRI \cite{min2018blind} predicts image quality by generating multiple pseudo-reference images obtained through further degradation of the distorted image and comparing their similarities. NFERM \cite{gu2014using} investigates image quality using the free energy principle.
Deep learning-based IQA methods have gained momentum with the advancement of deep neural networks. DBCNN \cite{zhang2018blind} consists of two streams of deep neural networks to address both synthetic and authentic distortions. HyperIQA \cite{su2020blindly} employs a self-adaptive hyper network to handle challenges arising from distortion diversity and content variation in IQA tasks. MUSIQ \cite{ke2021musiq}  utilizes a multi-scale image quality transformer to represent image quality at different levels of granularity. StairIQA \cite{sun2023blind} hierarchically integrates features extracted from intermediate layers to leverage low-level and high-level visual information.

% Most of the aforementioned IQA measures have demonstrated robust performance in predicting quality levels on traditional IQA databases, such as LIVE \cite{sheikh2006statistical}, TID2013 \cite{ponomarenko2015image}, CSIQ \cite{larson2010most}, and Kadid10K \cite{lin2019kadid}.
\subsubsection{Zero-shot NR-IQA}
Zero-shot IQA methods, also known as opinion-unaware methods, have emerged, which do not rely on training on subjective-rated quality assessment databases and can operate on unseen images. The earliest zero-shot NR-IQA methods are NIQE \cite{mittal2012making} and IL-NIQE \cite{zhang2015feature}. NIQE extracts handcrafted natural scene statistics (NSS) features from raw-resolution images and quantifies naturalness quality by computing the Multivariate Gaussian (MVG) distance to high-quality images. IL-NIQE enhances the feature set by incorporating additional quality-aware features, including gradient features, log Gabor filter responses, and color statistics.

\begin{figure*}
    \centering
    \includegraphics[width = 0.9\linewidth]{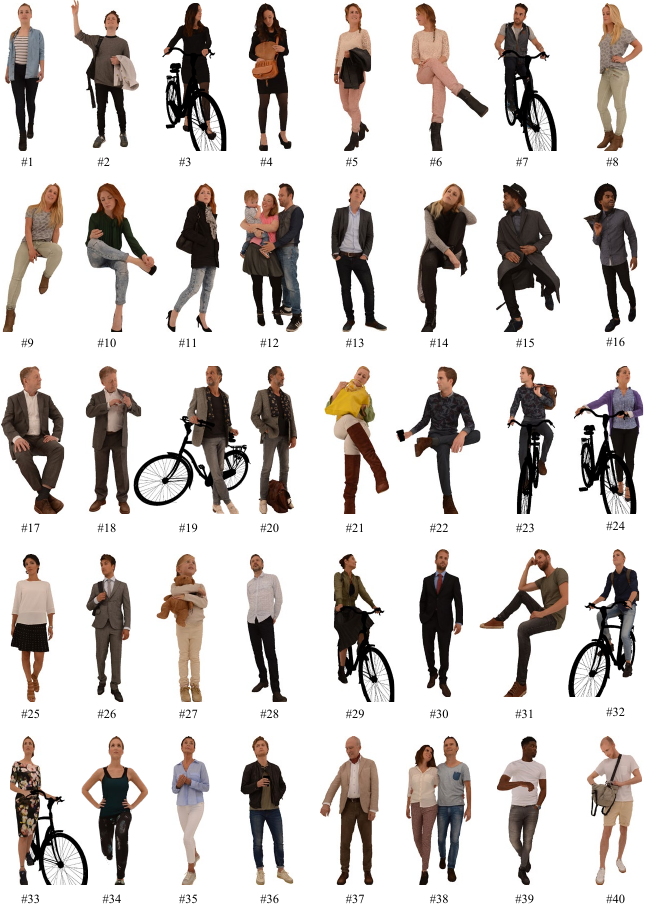}
    \caption{Overview of the reference 3D digital humans.}
    \label{fig:overview}
\end{figure*}

\begin{table}[!tp]
    \centering
    \caption{Number of the reference 3D digital humans' vertices and triangular faces.   }
    \begin{tabular}{c:c:c|c:c:c}
    \toprule
         ID & Vertices & Faces & ID & Vertices & Faces\\ \hline
         \#1 &21,185 &40,026 &\#21 &21,468 & 40,117 \\
         \#2 &21,681 &40,408 &\#22 &21,502 &40,462\\
         \#3 &22,558 &43,290 &\#23 &22,301 &42,778\\
         \#4 &21,585 &40,126 &\#24 &22,558 &43,290\\
         \#5 &21,274 &40,238 &\#25 &21,150 &40,098\\
         \#6 &21,326 &40,352 &\#26 &21,355 &40,203\\
         \#7 &22,302 &42,778 &\#27 &21,127 &40,462\\
         \#8 &21,434 &40,564 &\#28 &21,141 &40,116\\
         \#9 &21,336 &40,000 &\#29 &22,301 &42,778\\
         \#10 &21,211 &40,134 &\#30 &21,596 &40,746\\
         \#11 &22,532 &40,388 &\#31 &21,569 &40,488\\
         \#12 &21,608 &40,410 &\#32 &21,100 &40,046\\
         \#13 &20,791 &40,149 &\#33 &22,303 &42,780\\
         \#14 &21,287 &40,068 &\#34 &22,330 &42,778\\
         \#15 &21,657 &40,604 &\#35 &21,180 &40,284\\
         \#16 &21,339 &39,994 &\#36 &21,278 &40,236\\
         \#17 &21,159 &40,068 &\#37 &22,218 &40,010\\
         \#18 &21,378 &40,646 &\#38 &22,291 &39,996\\
         \#19 &21,362 &39,891 &\#39 &22,352 &39,998\\
         \#20 &22,301 &42,778 &\#40 &25,208 &40,052\\
    \bottomrule
    \end{tabular}
    \label{tab:number}
    \vspace{-0.3cm}
\end{table}

\section{Database Construction}
In this section, we mainly present the construction details of the proposed SJTU-H3D database, which includes reference collection, reference characterization, distortion generation, and subjective experiment. 

\begin{figure}
    \centering
    \includegraphics[width = \linewidth]{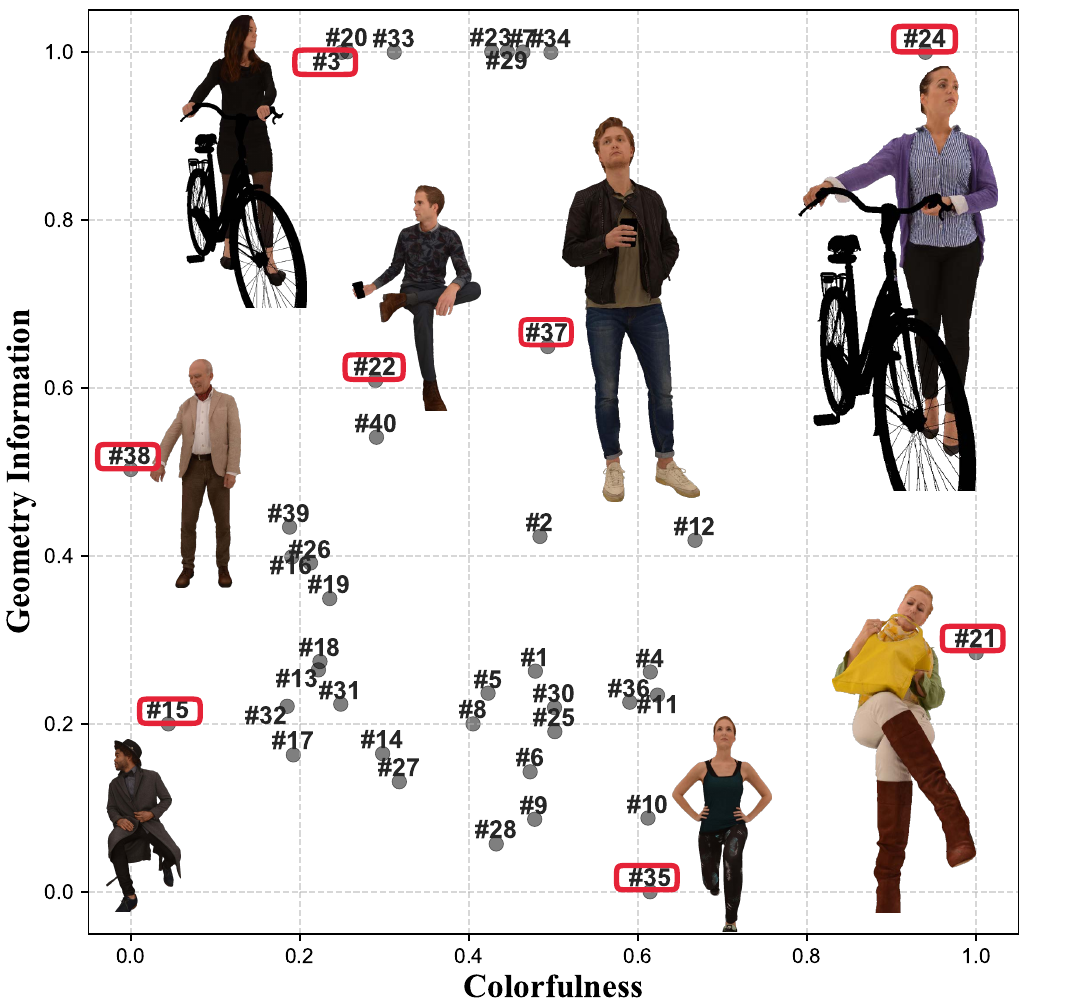}
    \caption{Geometry information and colorfulness for the reference 3D digital humans.}
    \label{fig:color_geo}
    \vspace{-0.4cm}
\end{figure}

\subsection{Reference Collection}
 In order to ensure the visual quality and content diversity of the reference 3D digital humans, a manual selection process is conducted to choose all reference digital humans from the HumanAlloy\footnote{https://humanalloy.com/}, a wonderful platform that provides high-quality 3D humans. A total of 40 digital humans are purchased and collected for this study. These digital humans are represented as textured meshes, with texture resolutions of 2048$\times$2048. Fig. \ref{fig:overview} illustrates the rendered projections of the selected digital humans, and Table \ref{tab:number} provides detailed information regarding the number of vertices and faces for each model.

\subsection{Reference Charaterization}
 The primary objective of our study is to curate a database that exhibits high diversity and generality while minimizing biases associated with the selection of source models. Therefore, we propose an approach to quantitatively characterize the geometry and color complexity since these two aspects are crucial for the visual quality of 3D digital humans. 

%collected 3D digital humans consist of geometry structure (.obj files) and textures (.jpg files)

\subsubsection{Geometry Information}
In the domain of image quality assessment (IQA), the analysis of spatial information often involves computing the standard deviation of the Sobel-filtered image. Motivated by this concept, we propose a novel approach to quantify the geometry information by utilizing the standard deviation of the dihedral angles in a mesh. The dihedral angle is a fundamental metric employed in computer graphics and geometric modeling to characterize the shape and curvature of meshes \cite{alexiou2018point,nr-svr}, thus drawing a parallel to the Sobel-filtering process in image analysis. It denotes the angle between two neighboring faces that share an edge within the mesh, providing valuable insights into the smoothness or sharpness of the surface. Specifically, the geometry information can be obtained as:
\begin{equation}
    GI = std(Mesh_{Dihedral}),
\end{equation}
where $GI$ represents the geometry information and $std(\cdot)$ stands for the standard deviation function.
By leveraging the standard deviation of dihedral angles, we aim to capture and assess the geometric characteristics of the mesh, enabling a more comprehensive evaluation of its structure and shape.

\subsubsection{Colorfulness}
To evaluate the color characteristics, we focus solely on the texture map. Following the common color calculation process \cite{fairchild2013color}, \cite{hasler2003measuring}, we first convert the texture from $RGB$ channels to $LAB$ channels and combine the standard deviation of $A$ and $B$ channels, which can be mathematically expressed as:
\begin{equation}
    CF = \sqrt{std(A)^2 + std(B)^2},
\end{equation}
where $CF$ represents the colorfulness measure, $A$ and $B$ denote the corresponding color channels of the texture. Similar colorfulness measures are also employed in many IQA works \cite{tian2011new,sheikh2006image,zhang2014fsim,yeganeh2012objective} as one of the metrics or features for assessing the quality of images.  

\subsubsection{Characterization Visualization}
We apply the extracted geometry information and colorfulness measure to the collection of 40 reference digital humans. The results are visualized in Fig. \ref{fig:color_geo}. The analysis demonstrates that the selected reference 3D digital humans exhibit a wide spectrum of geometry information and colorfulness. Notably, model \textbf{\#24} positioned in the top-right corner showcases intricate geometry details and vibrant colorfulness. In contrast, model \textbf{\#15} portrays simpler geometry information and relatively subdued colorfulness.
The proposed measures thoroughly capture the distinctiveness of 3D digital humans concerning their geometry and color characteristics. It is important to emphasize that these measures are directly computed from the underlying model files, thereby ensuring their stability and viewpoint invariance. 

\subsection{Distortion Generation}
To account for the common sources of distortion, we incorporate distortions arising from both the generation process and the transmission process. During the generation process, we consider geometry noise resulting from erroneous scanning procedures, as well as color noise introduced by cameras. Furthermore, compression and simplification techniques are widely employed during the transmission process. Hence, these factors are also taken into consideration in our assessment. By considering the full range of distortion sources, we aim to provide a comprehensive evaluation of the quality of 3D digital humans.

\begin{table}[!tp]\
    \centering
    \caption{Brief review of the parameters for distortion generation.}
    \begin{tabular}{c|l|l}
    \toprule
         Type & Parameter Description & Level \\ \hline
         GN & Geometry noise standard deviation & 0.05, 0.1, 0.15, 0.2\\  
         CN & Color noise standard deviation & 20, 40, 60, 80\\
         FS & Faces simplification rate & 0.4, 0.2, 0.1, 0.05 \\
         PC & Position compression parameter & 6, 7, 8, 9\\
         UMC & UV map compression parameter & 6, 7, 8, 9\\
         TD & Texture down-sampling rate & 2, 4, 8, 16 \\
         TC & Texture JPEG compression parameter & 3, 10, 15, 20 \\ 
    \bottomrule
    \end{tabular}
    \label{tab:distortion}
    \vspace{-0.3cm}
\end{table}

\begin{figure}[!tbp]
\centering

\subfigure[Ref]{
\centering
\includegraphics[width = 0.2 \linewidth]{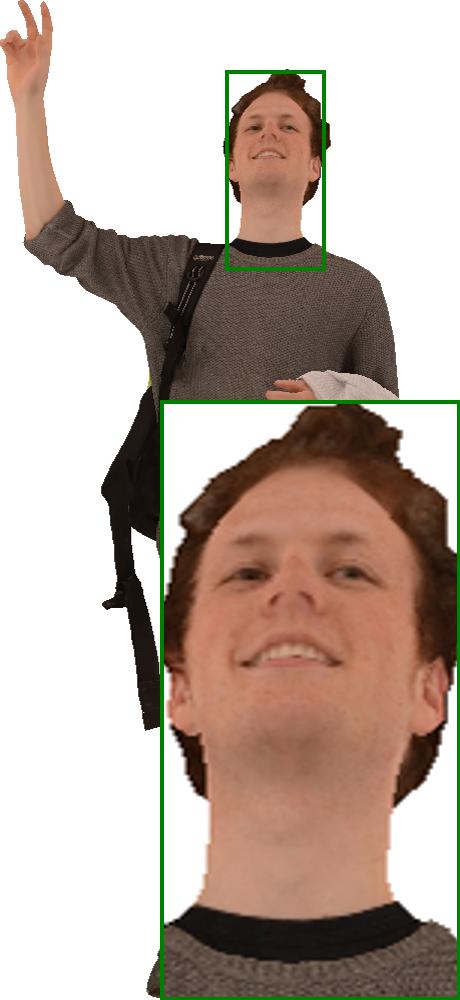}}
\subfigure[GN]{
\centering
\includegraphics[width = 0.2 \linewidth]{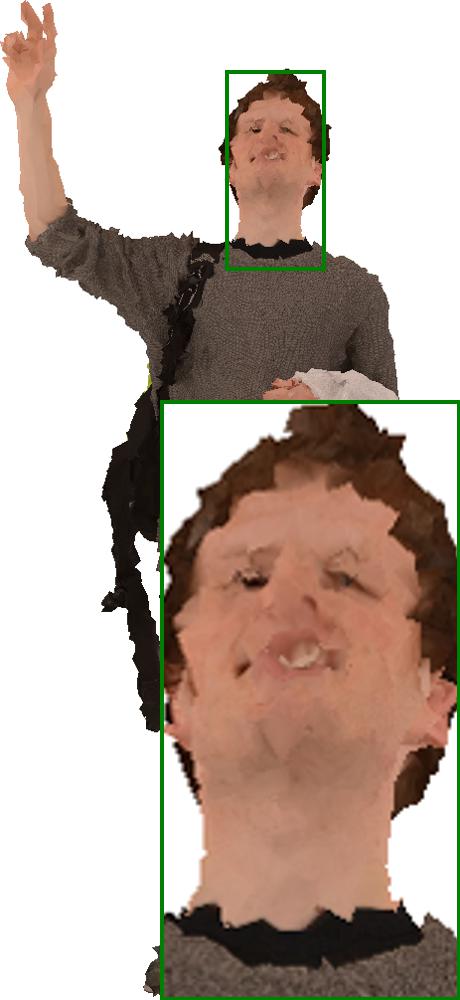}}
\subfigure[CN]{
\centering
\includegraphics[width = 0.2 \linewidth]{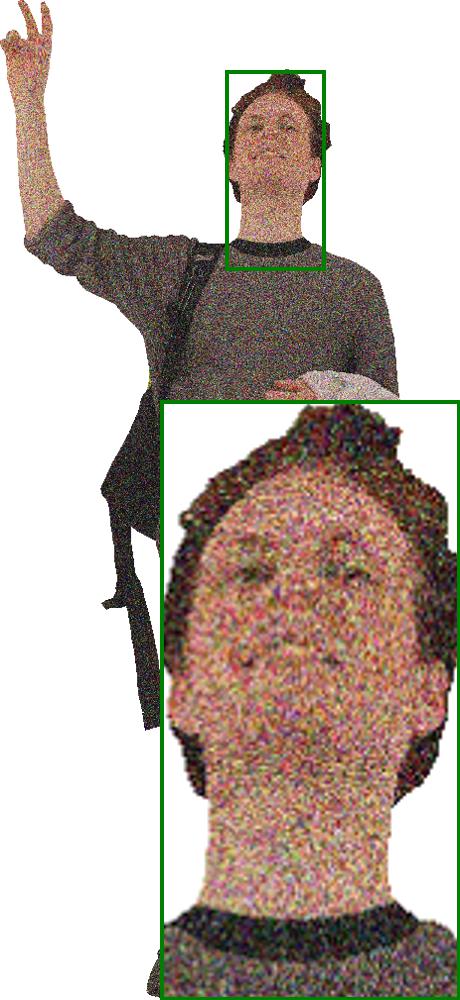}}
\subfigure[FS]{
\centering
\includegraphics[width = 0.2 \linewidth]{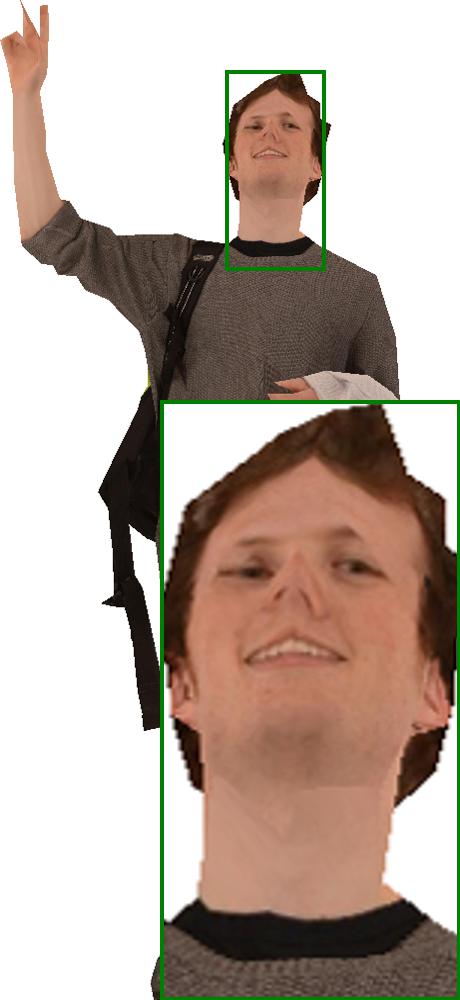}}

\subfigure[PC]{
\centering
\includegraphics[width = 0.2 \linewidth]{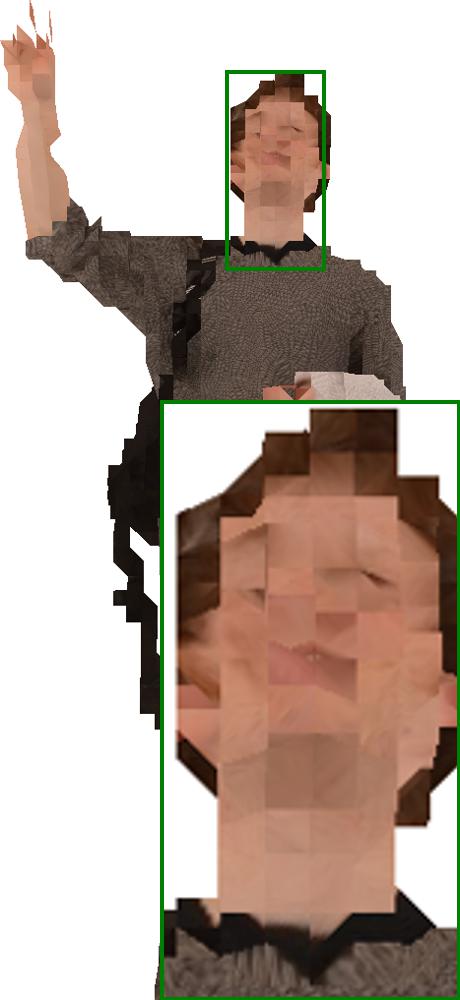}}
\subfigure[UMC]{
\centering
\includegraphics[width = 0.2 \linewidth]{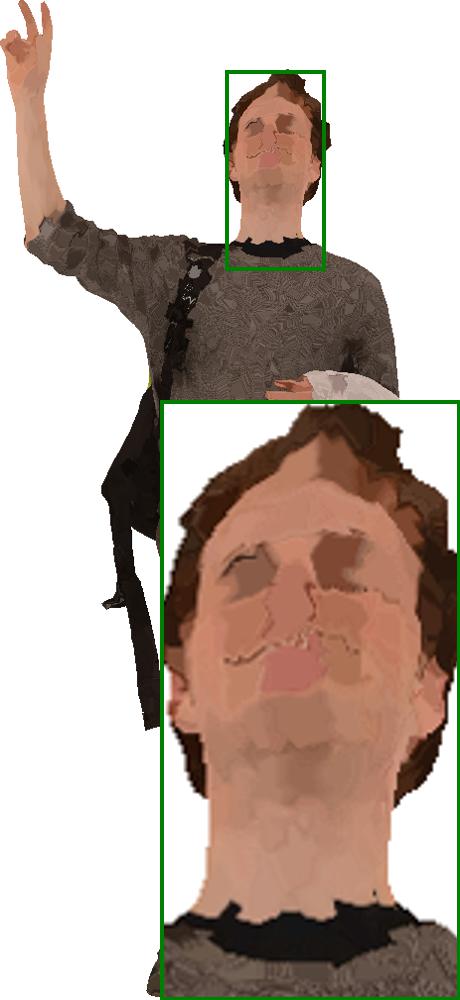}}
\subfigure[TD]{
\centering
\includegraphics[width = 0.2 \linewidth]{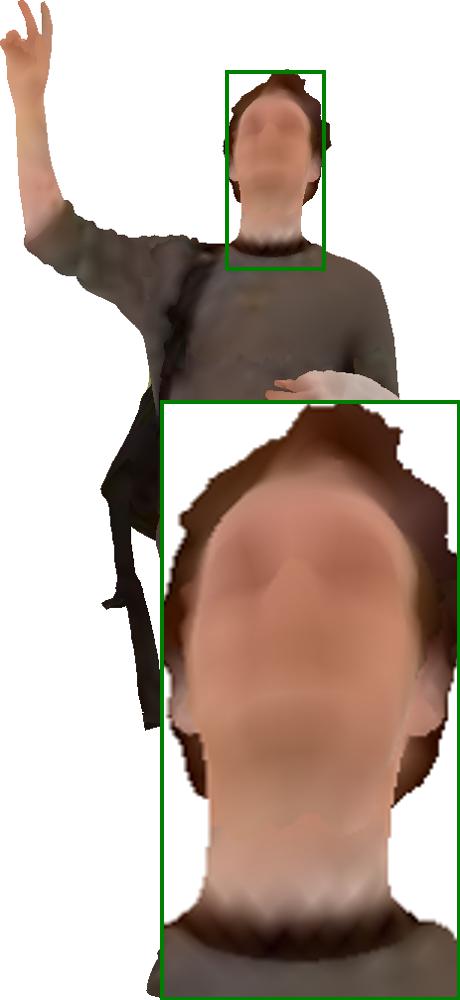}}
\subfigure[TC]{
\centering
\includegraphics[width = 0.2 \linewidth]{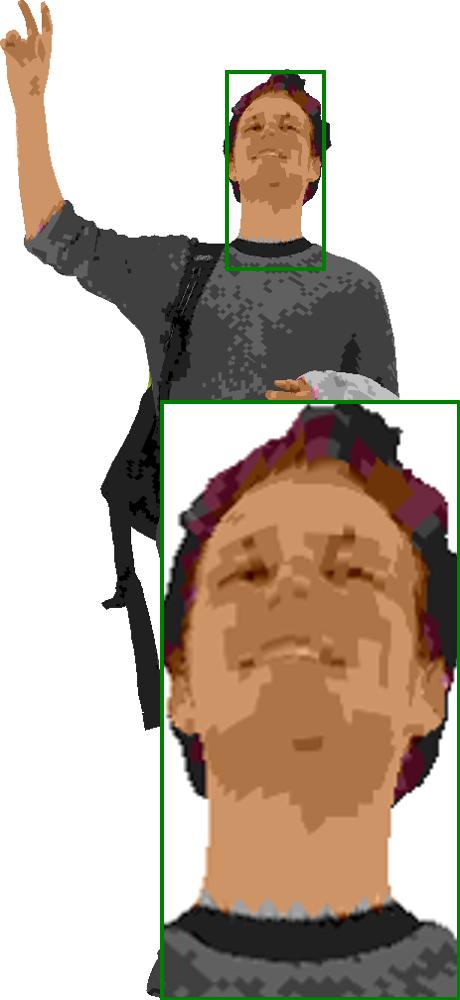}}
\caption{Examples of 3D digital humans with different types of distortions. Specifically, the presence of GN may introduce non-uniform irregular geometry artifacts to the model, while CN can give rise to noisy interference that negatively impacts visual quality. FS, on the other hand, can induce a simplification of the model's geometric representation, consequently leading to shape distortion. PC has the potential to generate blocky artifacts within the model. UMC can cause blurred texture misalignment, and TD can result in blurry and indistinct texture mapping. Additionally, TC may introduce color inaccuracies, while banding JPEG distortion can manifest as striped visual artifacts.}
\label{fig:distortion}
\vspace{-0.4cm}
\end{figure}

Therefore, to degrade the quality of the reference 3D digital humans, we apply seven types of distortions and the specific settings for each distortion type are listed in Table \ref{tab:distortion}. We manually select the distortion parameters to cover most visual quality range and the details are illustrated as follows:
\begin{itemize}
    \item Geometry Noise (GN): Gaussian noise with standard deviations $\sigma_g$ of 0.05, 0.1, 0.15, and 0.2 is added to the vertices' geometry coordinates of the digital humans.
    \item Color Noise (CN): Gaussian noise with standard deviations $\sigma_c$ of 20, 40, 60, and 80 is introduced to the texture maps.
    \item Face Simplification (FS): We utilize the simplification algorithm proposed in \cite{garland1998simplifying} to simplify the digital human. The simplification rate (number of faces remaining / number of original faces) is set to 0.4, 0.2, 0.1, and 0.05.
    \item Position Compression (PC): The Draco library\footnote{https://github.com/google/draco} is employed to quantize the position attributes of digital humans. The compression parameter $Q_p$ is varied as 6, 7, 8, and 9.
    \item UV Map Compression (UMC): Similarly, the Draco library is used to quantize the texture coordinate attributes with the compression parameter $Q_t$ set to 6, 7, 8, and 9.
    \item Texture Down-sampling (TD): The original texture maps with a resolution of 2048$\times$2048 are down-sampled to resolutions of 1024$\times$1024, 512$\times$512, 256$\times$256, and 128$\times$128.
    \item Texture Compression (TC): JPEG compression is employed to compress the texture maps. The quality levels are set to 3, 10, 15, and 20.
\end{itemize}

In all, a total of 40$\times$7$\times$4 = 1,120 distorted 3D digital humans are obtained. The distortion samples are exhibited in Fig. \ref{fig:distortion}.

\subsection{Subjective Experiment}

\subsubsection{Rendering Setting}

\begin{figure}
    \centering
    \subfigure[Rendering Process]{\includegraphics[width = 0.7\linewidth]{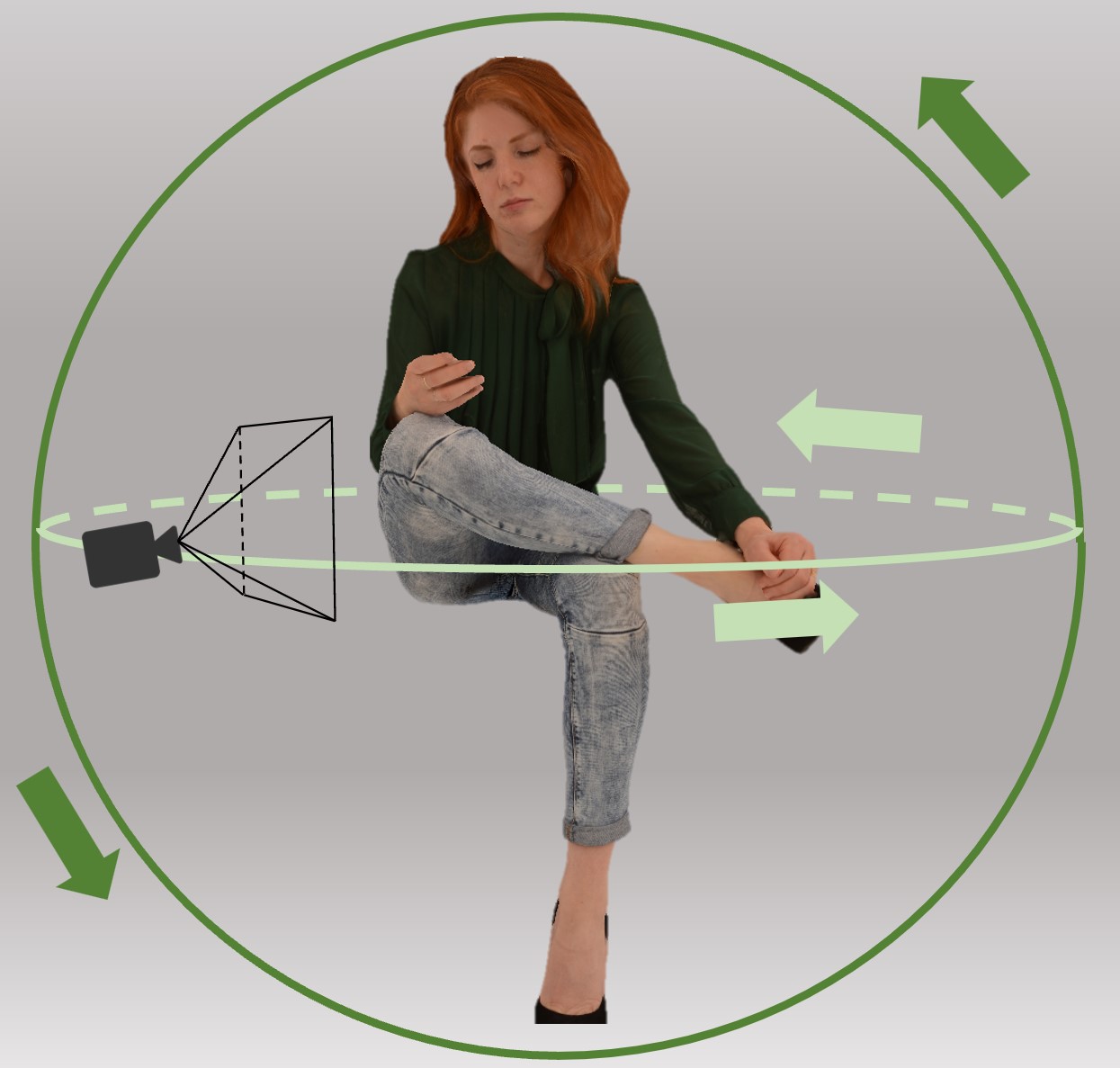}}
     \subfigure[Rating Interface]{\includegraphics[width = 0.7 \linewidth]{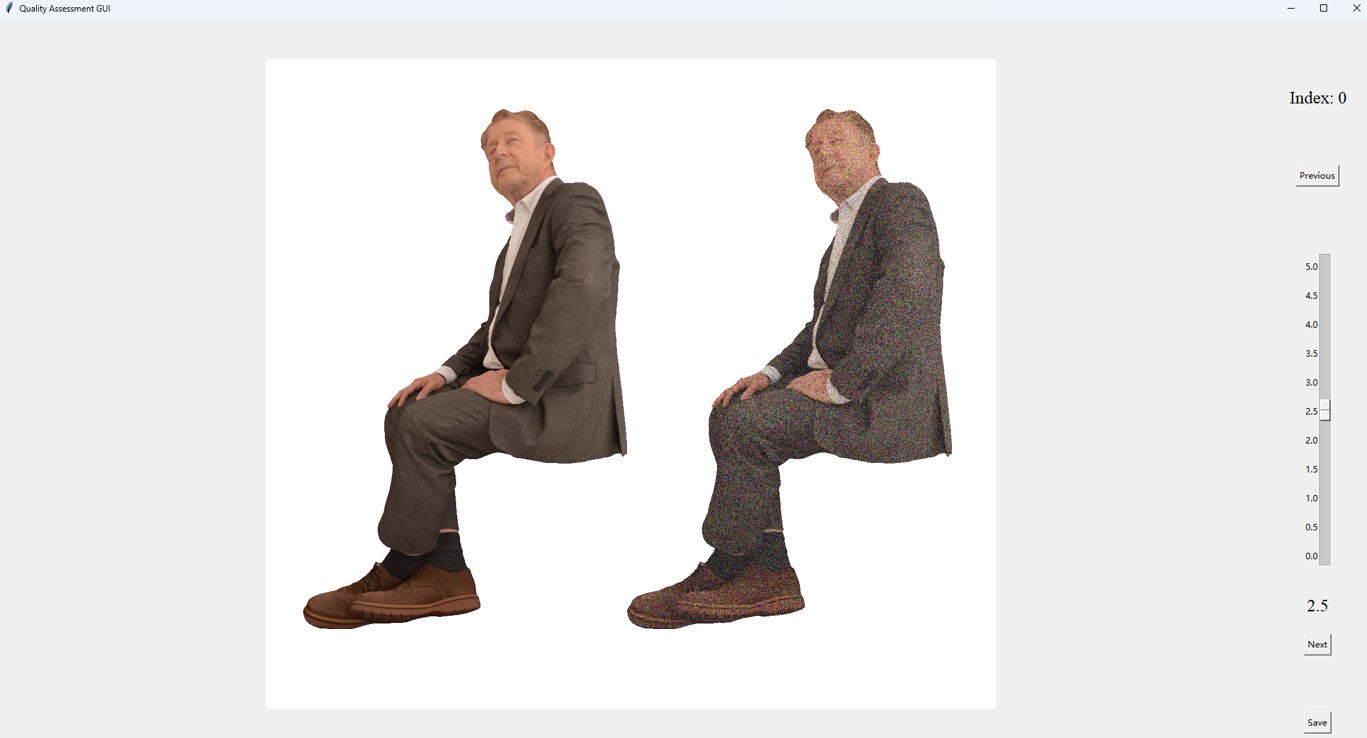}}
     \caption{Inllustration of the rendering process and rating interface.}
    \label{fig:rendering}
    \vspace{-0.4cm}
\end{figure}

In accordance with the recommended procedure outlined in \cite{yang2020predicting,liu2022perceptual}, passive watching is chosen over interactive watching for the subjective experiment to mitigate potential viewing bias. The 3D digital humans are rendered into video sequences for exhibition purposes. The \textit{open3d} library is utilized to generate the projections \cite{Zhou2018}. The rendering window is configured with a resolution of $1080 \times 1920$. To capture the video frames, a horizontal and a vertical circle are employed as the predefined camera paths. Each 3D digital human is captured at one frame every $3$ degree, resulting in a total of $240$ frames ($360 \times 2 \div 3$). These frames are then compiled into an 8-second video with a framerate of 30 frames per second. This approach ensures that the viewers can effectively perceive the significant quality information. The rendering process is depicted in Fig. \ref{fig:rendering}.

\subsubsection{Experiment Process}
A total of 40 human subjects, comprising 20 males and 20 females, are recruited to participate in the subjective experiment. Prior to the experiment, a training session is conducted, wherein additional videos generated using the same aforementioned process are presented to familiarize the subjects with the tasks. The rating process takes place within a well-controlled laboratory environment, maintaining a normal level of illumination. The viewers are seated at a distance of twice the screen height. The videos are displayed on an iMac monitor capable of supporting resolutions up to $4096 \times 2304$. The order of video presentations is randomized. To facilitate the evaluation process, a double stimuli strategy is employed, where the reference and distorted videos are simultaneously displayed on the screen. The rating interface is excited in Fig. \ref{fig:rendering} and the quality score ranges from 0 to 5. In order to mitigate viewer fatigue, the entire experiment is divided into 20 sessions, with each session featuring 56 digital humans. Ultimately, a total of 44,800 subjective ratings ($1,120 \times 40$) are collected.

\begin{figure}
    \centering
    \subfigure[Overall MOS Distribution]{\includegraphics[width = 0.8\linewidth]{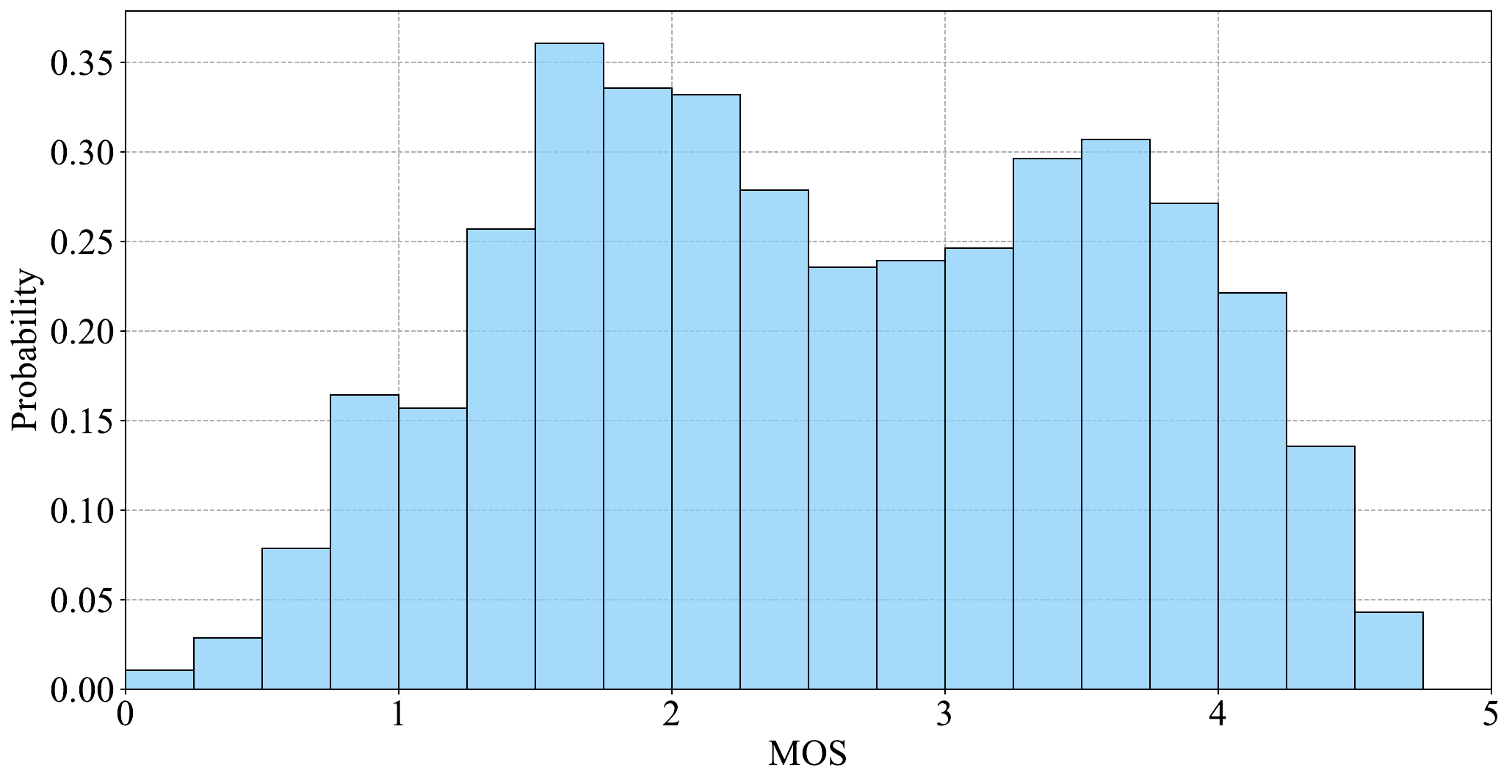}}
    \subfigure[Distortion-specific MOS Distribution ]{\includegraphics[width = 0.8\linewidth]{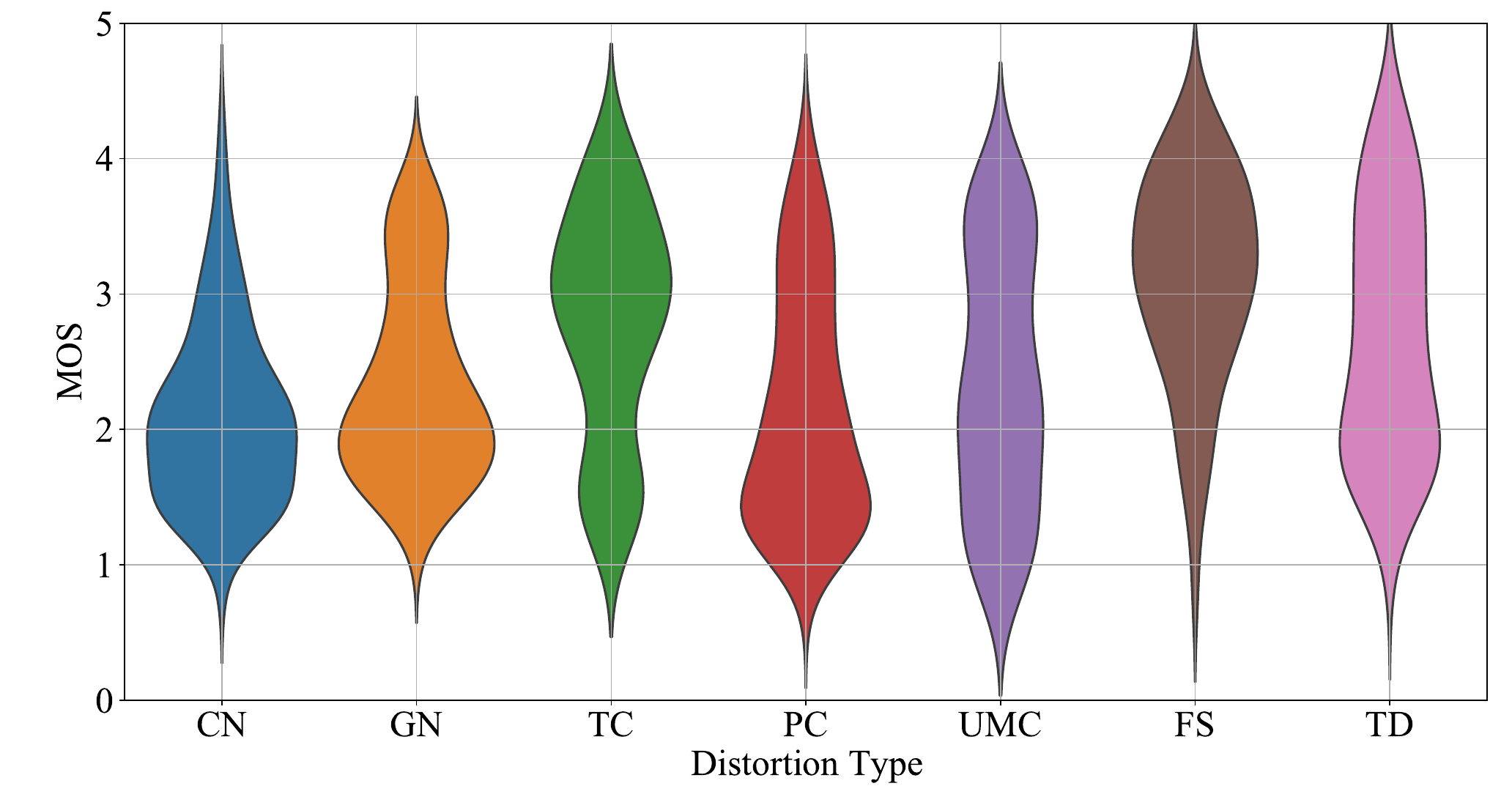}}
    \caption{Illustration of the probability distribution of MOS and the probability distributions corresponding to different distortion types. }
    \label{fig:mos}
    \vspace{-0.4cm}
\end{figure}

\subsubsection{Subjective Data Analysis}
After the subjective experiment, we calculate the z-scores from the raw ratings as follows:
\begin{equation}
z_{i j}=\frac{r_{i j}-\mu_{i}}{\sigma_{i}},
\end{equation}
where $\mu_{i}=\frac{1}{N_{i}} \sum_{j=1}^{N_{i}} r_{i j}$, $\sigma_{i}=\sqrt{\frac{1}{N_{i}-1} \sum_{j=1}^{N_{i}}\left(r_{i j}-\mu_{i}\right)}$, and $N_i$ is the number of digital humans judged by subject $i$.
In accordance with the ITU-R BT.500-13 \cite{bt2002methodology} standard, ratings from unreliable subjects are excluded from the analysis.
The corresponding z-scores are linearly rescaled to the range of $[0,5]$. Finally, the mean opinion scores (MOSs) are computed by averaging the rescaled z-scores. 

Fig. \ref{fig:mos} illustrates the distribution of MOSs and the corresponding probability distributions for different distortion types. Interestingly, the probability distributions reveal that visual quality is less sensitive to varying levels of FS distortions compared to other distortion types. Even when reducing the face numbers to a ratio of 0.05 (only about 2k faces are preserved), the visual quality score remains higher than other distortions with similar levels. This observation indicates that visual quality is relatively resilient to FS distortions, implying that the reduction in face complexity may not significantly impact the perceived quality.

\begin{figure*}
    \centering
    \includegraphics[width = 0.8\linewidth]{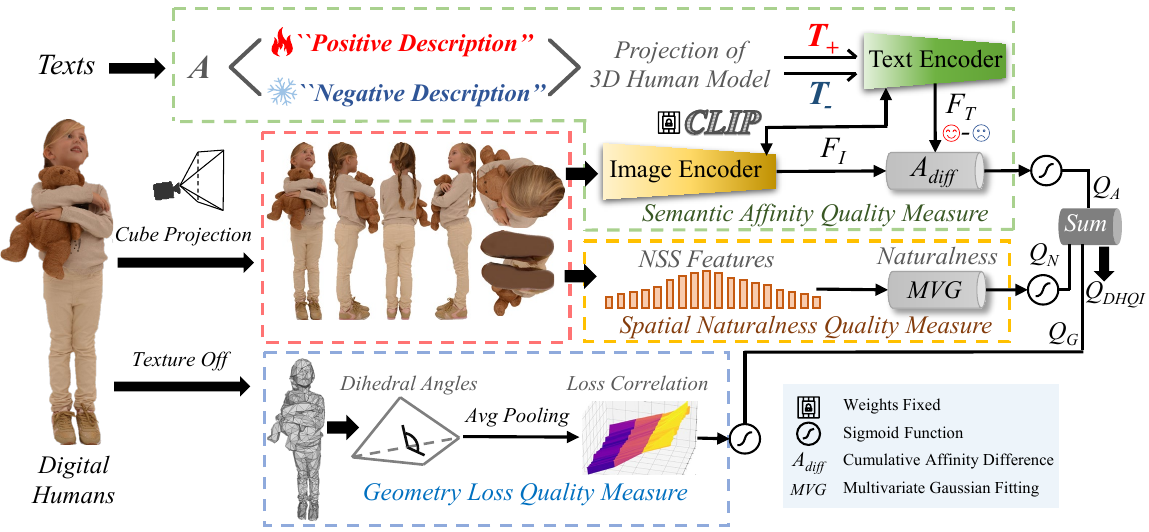}
    \caption{The framework of the proposed method.}
    \label{fig:framework}
    \vspace{-0.4cm}
\end{figure*}

\section{Proposed Method}
In this section, we introduce the three indexes that make up the whole proposed digital human quality index (DHQI), which includes the text-prompted semantic affinity quality measure, spatial naturalness quality measure, and geometry loss quality measure. These three indexes are then aligned and aggregated into the proposed DHQI quality index. The framework is exhibited in Fig. \ref{fig:framework}.

\subsection{Pre-processing}
\label{sec:pre}
We acquire the cube-like projection set of the given digital human as follows:
\begin{equation}
\begin{aligned}
     &\mathcal{P}  = \psi(\mathcal{DH}), \\
    \mathcal{P}  = \{&P_{k}|k =1,\cdots, 6\},
\end{aligned}
\end{equation}
where $\mathcal{P}$ represents the set of the 6 rendered projections and $\psi(\cdot)$ stands for the rendering process. 
Such rendering process has been employed in the popular point cloud compression standard MPEG VPCC \cite{graziosi2020overview} and many other 3DQA works \cite{yang2020predicting,zhang2022treating}. The projections are utilized as the input information for the text-prompted semantic affinity and spatial naturalness measure.

\subsection{Text-prompted Semantic Affinity Quality Measure}
To assess the perception of quality related to semantic content, specifically evaluating the quality of contents and the ability to discern semantic distortions, we design the text-prompted semantic affinity quality measure. Inspired by CLIP \cite{radford2021learning}-based quality assessment tasks \cite{wu2023towards,wang2022exploring}, we hold the hypothesis that the projections of the high-quality digital humans should have higher affinity with {\bf\color{red} \textit{positive}} quality-related descriptions ($e.g.$ \textit{good}, \textit{perfect}) and lower affinity with {\bf\color{blue} \textit{negative}} quality-related descriptions ($e.g.$ \textit{bad}, \textit{distorted}). 

\subsubsection{Text Prompt Format}
\label{sec:text}
In accordance with the official recommendation provided by CLIP \cite{radford2021learning} and drawing from established practices, our text prompts are designed as a concatenation of three components: a prefix, a description, and a suffix. To be more precise, the text prompt $T$ corresponding to the raw description $D$ is defined as:
\begin{equation}
    T = ``a" + D + ``projection \!\!\!\quad of \!\!\!\quad 3d \!\!\!\quad human \!\!\!\quad model",
    \label{equ:text}
\end{equation}
where the suffix ``\textit{projection of 3d human model}" is specifically designed to fit the task of DHQA. This carefully chosen suffix can encourage the CLIP model to prioritize and focus its attention on the detection and evaluation of content-aware distortions that may arise in the context of 3D digital humans.

\subsubsection{Description Selection}
We have identified descriptions pertaining to quality assessment that encompass broad evaluation aspects to ensure robustness. In this study, the general quality-related descriptions employed comprise the contrasting pairs of \textit{high quality} $\leftrightarrow$ \textit{low quality}, \textit{good} $\leftrightarrow$ \textit{bad}, and \textit{perfect} $\leftrightarrow$ \textit{distorted}. The utilization of the \textit{high quality} $\leftrightarrow$ \textit{low quality} as well as the \textit{good} $\leftrightarrow$ \textit{bad} text pair assists in directing the attention of the CLIP model towards general subjective impressions. Conversely, the \textit{perfect} $\leftrightarrow$ \textit{distorted} pair compels the CLIP model to prioritize the existence of distortions.

% Taking into consideration the characteristics of distortions, we have included distortion-specific descriptions such as \textit{sharp} $\leftrightarrow$ \textit{blurry}, \textit{rough} $\leftrightarrow$ \textit{smooth}, and \textit{complex} $\leftrightarrow$ \textit{simple}.

\subsubsection{Affinity Difference Computation}
Given the input image $I$ and text $T$, the senmantic affinity can be calculated with the assistance of CLIP as:
\begin{equation}
    \begin{aligned}
        F_I &= E_I(I), \quad F_T = E_T(T), \\
        &A(I,T) = \frac{F_I \cdot F_T'}{\lVert F_I \rVert \lVert F_T \rVert},
    \end{aligned}
\end{equation}
where $E_I$ and $E_T$ stand for the image and text encoders of CLIP, $F_I$ and $F_T$ represent the CLIP-encoded features, and $A(I,T)$ indicates the affinity between the input image and text.
Afterward, the computation of zero-shot quality affinity can be derived from the aforementioned selected descriptions by calculating the disparity between the probabilities assigned to positive and negative textual inputs:
\begin{equation}
\begin{aligned}
\mathcal{A}(\mathcal{P},&T) = \frac{1}{6}\sum_{K=1}^6 A(P_{k},T),\\
\mathcal{A}_{diff}(\mathcal{P},T_+, T_-) &= \frac{1}{N_T}\sum_{i=1}^{N_T}(\mathcal{A}(\mathcal{P},T_+^i) - \mathcal{A}(\mathcal{P},T_-^i)),
\end{aligned}
\end{equation}
where the averaged affinity to the given text $T$, denoted by $\mathcal{A}(\mathcal{P}, T)$, is calculated by CLIP across the six projections $\mathcal{P}$. In this context, $T_+^i$ and $T_-^i$ refer to the positive and negative text descriptions, respectively, from the $i$-th text pair. The variable $N_T$ represents the total number of text pairs. Furthermore, $\mathcal{A}_{diff}$ signifies the cumulative difference between the averaged positive and negative affinity.

The sigmoid remapping technique is then used to map the raw difference scores $\mathcal{A}_{diff}$ obtained from perceptual quality evaluation into a range of [0, 1]. This remapping is done based on the guidance provided by the Video Quality Experts Group (VQEG) \cite{video2000final}.
The purpose of sigmoid remapping is to transform the raw difference scores into a perceptually meaningful range that is easier to interpret, and the final text-prompted semantic affinity quality score can be derived as:
\begin{equation}
    Q_A = \frac{1}{1 + e^{-\mathcal{A}_{diff}}},
\end{equation}

\subsection{Spatial Naturalness Quality Measure}
\label{sec:niqe}
Apart from evaluating semantic affinity, we incorporate the use of NIQE (Naturalness Image Quality Evaluator \cite{mittal2012making}) as a blind quality evaluator to assess the spatial naturalness of the digital humans. The purpose of employing NIQE is to identify and quantify common low-level distortions encountered in practical digital humans, including Gaussian noise, blur, and JPEG compression artifacts. By incorporating NIQE alongside semantic affinity evaluation, we aim to complement the assessment of high-level information with an evaluation of low-level technical quality.

The NIQE index operates by quantifying the disparity between the characteristics of the input image features and the anticipated distribution of features observed in "high-quality" images, which are derived from a diverse set of pristine natural images. Since the raw NIQE scores and the raw affinity difference scores are on different scales, it is necessary to normalize the NIQE scores to facilitate meaningful comparison. To achieve this, we divide the NIQE scores by a constant value, denoted as $c_1$, which effectively restricts the majority of NIQE scores to the range of [0,1]. Consequently, the spatial naturalness quality measure can be computed as follows:

\begin{equation}
\begin{aligned}
\mathcal{N}(\mathcal{P}) &= \frac{1}{6}\sum_{K=1}^6 N(P_{k}),\\
Q_N &= \frac{-1}{1 + e^{-\mathcal{N}/c_1}}
\end{aligned}
\end{equation}
where $N(P_{k})$ denotes the NIQE value for the $k$-th projection, $\mathcal{N}(\mathcal{P})$ represents the average NIQE value across the 6 projections, and $Q_N$ stands for the spatial naturalness quality measure. It's worth noting that the NIQE scores are inversely correlated with quality and the negative sign is incorporated into the sigmoid function, allowing for a consistent interpretation and alignment of the NIQE scores with the quality evaluation framework.

\begin{figure}
    \centering
    \subfigure[Position Compression]{\centering \includegraphics[width = 0.45\linewidth]{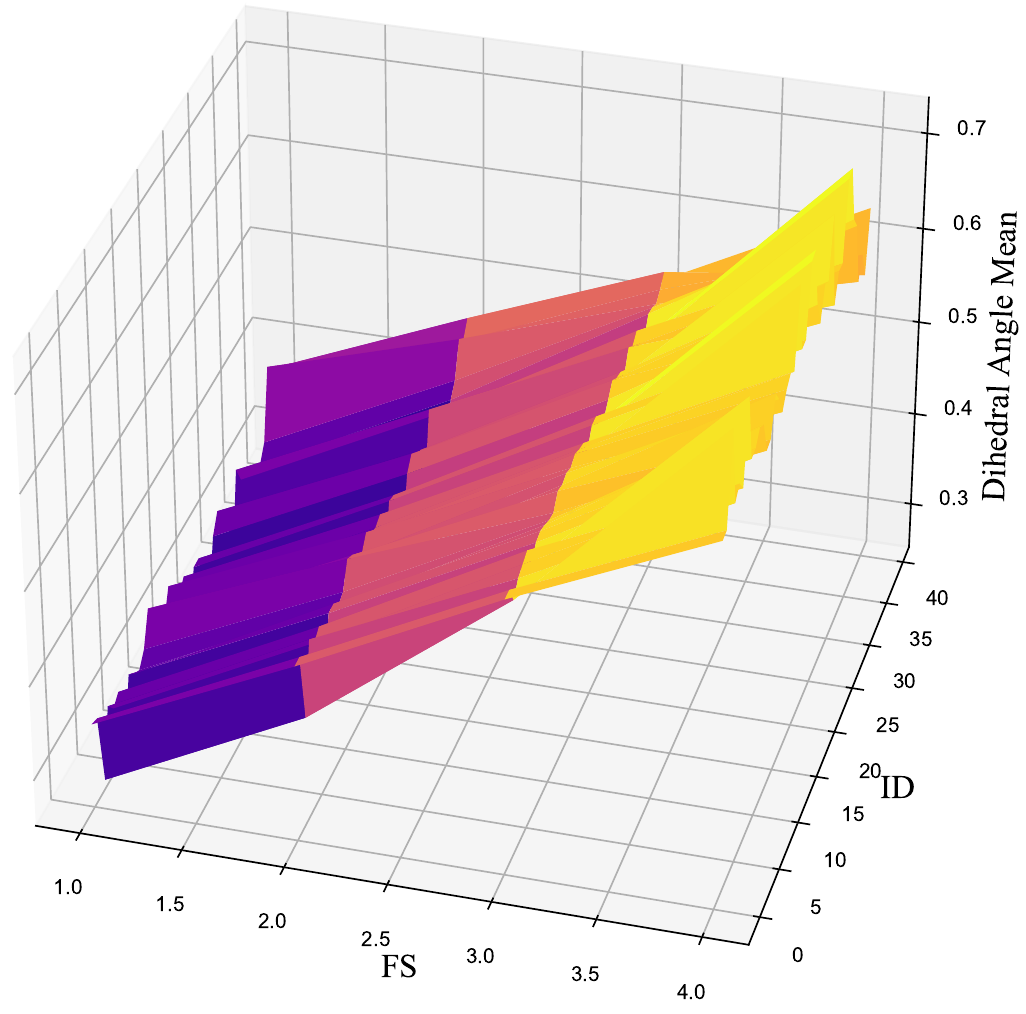}}
    \subfigure[Face Simplification]{\centering \includegraphics[width = 0.45\linewidth]{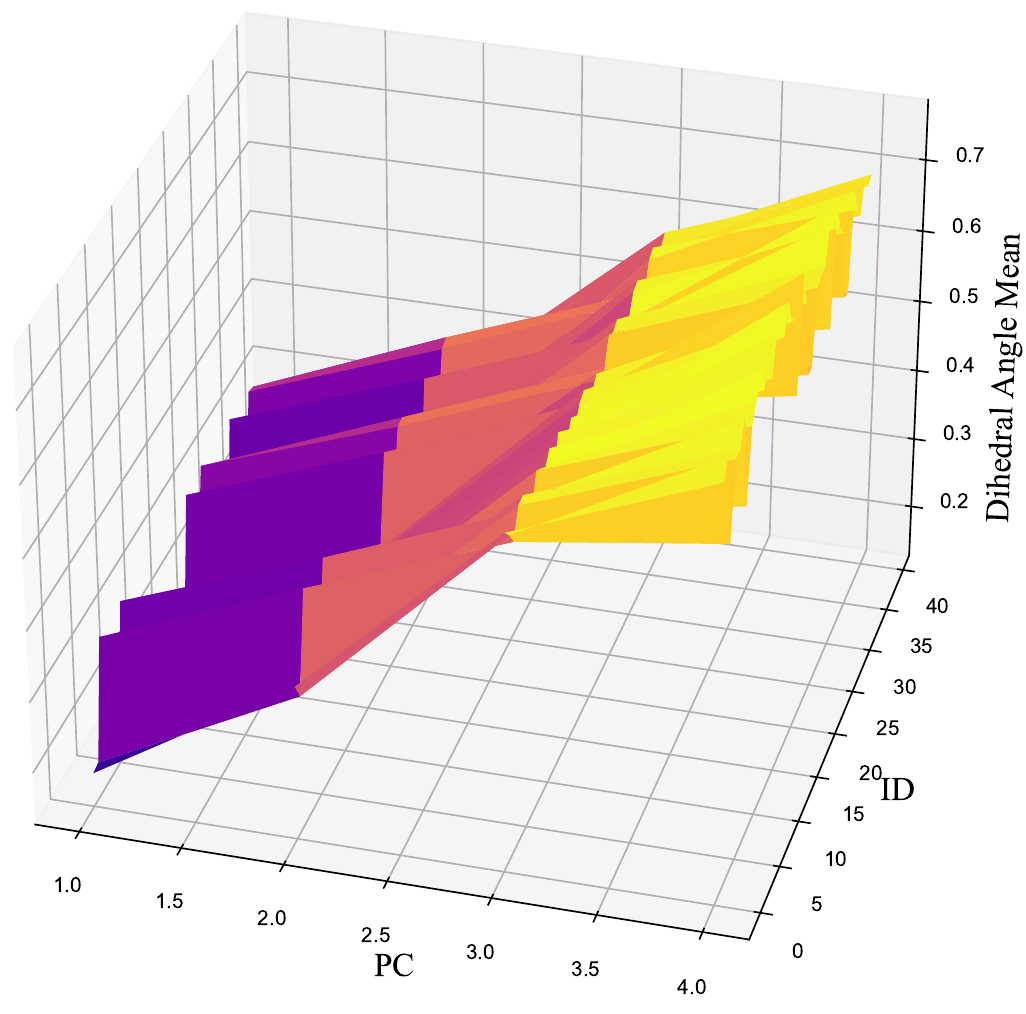}}
    \caption{ Illustration of the increasing tendency of mean dihedral angles in relation to compression/simplification distortion levels in digital humans. The x-axis denotes the levels of distortion, while the y-axis represents the IDs assigned to the digital humans. The z-axis corresponds to the mean values of dihedral angles observed in the distorted digital humans. }
    \label{fig:dihedral}
    \vspace{-0.4cm}
\end{figure}

\subsection{Geometry Loss Measure}
The aforementioned measures are applied to projections, specifically the image modality. In order to enhance the model's understanding of digital humans, it is proposed to directly extract features from the mesh modality to capture the loss in geometry with respect to visual quality.

\subsubsection{Descriptor Selection} 
Various geometry attributes have been utilized to describe the quality-related geometric characteristics of meshes \cite{zhang2021mesh}, including curvature, dihedral angle, face angle, face area, etc. For the purpose of preserving stability and improving the robustness of the proposed zero-shot method, the dihedral angle is selected as the geometry descriptor for the following reasons: a) Extensive evidence supports the effectiveness of the dihedral angle in describing geometric features relevant to visual quality \cite{alexiou2018point,nr-svr,dame,zhang2021mesh}. b) Unlike other geometry attributes, the dihedral angle is invariant to scale. Its values are confined within the range of [0, $\pi$], thereby contributing to its robustness.
The dihedral angle is the angle between two adjacent faces, which can be calculated as the dot product of corresponding normal vectors:
\begin{equation}
    \Theta = \{\frac{\theta_{j}}{\pi} | cos(\theta_{j}) = \frac{{n}_{j1} \cdot {n}_{j2}}{\lVert {n}_{j1} \rVert \lVert {n}_{j2} \rVert} \},
\end{equation}
where $\frac{\theta_{j}}{\pi}$ indicates the scaled dihedral angle corresponding to the $j$-th edge of the mesh, $\Theta$ indicates the set of the scaled dihedral angle values, ${n}_{j1}$ and ${n}_{j2}$ stand for the normal vectors of the two adjacent faces whose co-edge is the $j$-th edge.

\subsubsection{Quality Correlation with Dihedral Angle}
Lossy mesh compression and simplification techniques can potentially diminish a mesh's structural details, resulting in a smoother and simpler surface representation. In such cases, the faces comprising the smoother and simpler surface tend to exhibit dihedral angles that approach $\pi$, leading to an inherent inclination for larger dihedral angles. To substantiate this observation, we present the tendencies of the mean values of the dihedral angles in Fig. \ref{fig:dihedral}, from which we can find a consistent upward trend in dihedral angle means as compression/simplification levels increase. Therefore, the mean values of the dihedral angle can be generally taken as an indicator of geometry detail loss caused by compression/simplification. Then geometry loss quality measure can be calculated as:
\begin{equation}
    Q_G = \frac{-1}{1 + e^{-\Bar{\Theta}}},
\end{equation}
where $Q_G$ represents the geometry loss quality measure, $\Bar{\Theta}$ indicates the mean value of the dihedral angles, and the negative sign is added to the sigmoid function due to the positive correlation between the dihedral angles' mean values and compression/simplification levels. 

\subsection{Quality Measure Aggregation}
In order to develop a reliable zero-shot perceptual quality index, we adopt a direct aggregation approach wherein we sum up the scale-aligned scores of various indices without performing any fine-tuning processes. Considering that the $Q_A$, $Q_N$, and $Q_G$ have undergone sigmoid rescaling, all three measures are bounded within the range of [0, 1]. Consequently, we define the comprehensive unified DHQI (digital human quality index) as follows:
\begin{equation}
    Q_{DHQI} = Q_A + Q_N + Q_G,
\end{equation}
where $Q_{DHQI}$ indicates the final quality values for the digital humans.

\section{Experiment}

\begin{table*}[!htp]
    \centering
    \caption{Benchmark Performance on the SJTU-H3D and DHHQA databases. The methods marked with * are fixed with prestrained weights. The best and second performances for the zero-shot and supervised methods are marked in {\bf\textcolor{red}{RED}} and {\bf\textcolor{blue}{BLUE}} respectively.}
    \begin{tabular}{l:l:c:l|cccc|ccccc}
    \toprule
       \multirow{2}{*}{Type}  &  \multirow{2}{*}{Ref} & \multirow{2}{*}{Index} & \multirow{2}{*}{Method} & \multicolumn{4}{c|}{SJTU-H3D} & \multicolumn{4}{c}{DHHQA}\\ \cline{5-12}
         & && & SRCC$\uparrow$  &  PLCC$\uparrow$ & KRCC$\uparrow$ & RMSE$\downarrow$ & SRCC$\uparrow$  &  PLCC$\uparrow$ & KRCC$\uparrow$ & RMSE$\downarrow$ \\ \hline
        \multirow{13}{20pt}{\bf{Zero-shot}} & \multirow{8}{20pt}{FR}&A& PSNR  & 0.5139 & 0.4974 & 0.31706 & 0.9441 & 0.8347 & \bf\textcolor{blue}{0.8371} & 0.6405 & \bf\textcolor{blue}{11.5822} \\
        & &B&  SSIM (TIP, 2004) \cite{ssim} & \bf\textcolor{blue}{0.7336} & \bf\textcolor{blue}{0.6888} & \bf\textcolor{blue}{0.5416} & \bf\textcolor{blue}{0.8626} & 0.7355 & 0.7253 & 0.5388 & 14.5221\\
        & &C& MS-SSIM (ACSSC, 2004) \cite{wang2003multiscale} & 0.2417 & 0.2776 & 0.1822 & 1.0150 & \bf\textcolor{red}{0.8557} & \bf\textcolor{red}{0.8396} & \bf\textcolor{red}{0.6653} & \bf\textcolor{red}{11.4953}\\
        & &D& GMSD (TIP, 2013) \cite{xue2013gradient} & 0.2574 & 0.3538 & 0.1855 & 0.9833 & \bf\textcolor{blue}{0.8411} & 0.8350 & \bf\textcolor{blue}{0.6534} & 11.6441\\ 
        & &E& PSNR$_{p2po}$ (MPEG, 2017) \cite{tian2017evaluation} & 0.2636 & 0.2680 & 0.2154 & 1.0134 & 0.2891 & 0.2916 & 0.2359 & 21.0813\\
        & &F& PSNR$_{p2pl}$ (MPEG, 2017) \cite{tian2017updates}  & 0.2101 & 0.2114 & 0.1686 & 1.0244 & 0.2698 & 0.2961 & 0.2250 & 21.0520\\
        & &G& PSNR$_{yuv}$ (MPEG, 2017) \cite{mekuria2016design} & 0.5247 & 0.5638 & 0.4141 & 0.9199 & 0.1761 & 0.2272 & 0.1369 & 21.4299\\
        & &H& G-LPIPS* (TOG, 2022) \cite{nehme2022textured} & 0.6930 & 0.6112 & 0.5343 & 0.7966 & 0.8389 & 0.8069 & 0.6446 & 12.0051\\ \cdashline{2-12}
        & \multirow{5}{20pt}{NR} &I& CPBD (TIP, 2011) \cite{narvekar2011no}  & 0.3643 & 0.4936 & 0.2563 & 0.9102 & 0.2621 & 0.2599 & 0.1718 & 20.6701 \\
        &&J& BRISQUE* (TIP, 2012) \cite{mittal2012brisque}  & 0.3312 & 0.3623 & 0.2417 & 0.9799 & 0.4552 & 0.5158 & 0.3011 & 18.1009 \\
        &&K& NIQE (TIP, 2012) \cite{mittal2012making} & 0.4506 & 0.5546 & 0.3266 & 0.8665 & 0.4232 & 0.4410 & 0.2784 & 18.9814 \\
        & &L& IL-NIQE (TIP, 2015) \cite{zhang2015feature} & 0.3683 & 0.5239 & 0.2595 & 0.8879 & 0.5419 & 0.6232 & 0.3825 & 17.4119\\
        & &M& \textbf{DHQI (Proposed)} & \bf\textcolor{red}{0.8296} & \bf\textcolor{red}{0.8276} & \bf\textcolor{red}{0.6310} & \bf\textcolor{red}{0.5689} & \bf{0.7883} & \bf{0.8070} & \bf{0.5955} & \bf{12.4795}  \\  \hline
        \multirow{8}{20pt}{\bf{Super-vised}} 
        % & \multirow{2}{20pt}{FR} & LPIPS  & 0.7853 & 0.7821 & 0.5869 & 0.6534 & 0.8935 & 0.8881 & 0.7085 & 9.7194 \\
        % & & SSIM (TIP, 2004) \cite{ssim} & 0.1111 & 0.1111 & 0.1111 & 0.1111 & 0.1111 & 0.1111 & 0.1111 & 0.1111\\ \cdashline{2-11}
        & \multirow{8}{20pt}{NR} &N& BRISQUE (TIP,2012) \cite{mittal2012brisque} & 0.6029 & 0.6242 & 0.4317 & 0.8567 & 0.6008 & 0.5958 & 0.4214 & 17.3699  \\
        & &O& NFERM (TMM, 2014) \cite{gu2014using} & 0.4091 & 0.4452 & 0.2895 & 0.9535 & 0.6414 & 0.6876 & 0.4610 & 15.4528\\ 
        & &P& BMPRI (TBC, 2018) \cite{min2018blind} & 0.4697 & 0.5314 & 0.3352 & 0.9049 & 0.8011 & 0.8065 & 0.6106 & 13.0748\\ 
        % & & NFSDM & 0.2015 & 0.1956 & 0.1364 & 1.0547 & 0.5921 & 0.6598 & 0.4214 & 16.1233\\
        & &Q& DBCNN (TCSVT, 2018) \cite{zhang2018blind} & 0.7428 & 0.7408 & 0.5397 & 0.7030 & 0.7575 & 0.8238 & 0.5723 & 12.0087\\
        & &R& HyperIQA (CVPR, 2020) \cite{su2020blindly} & 0.7986 & 0.7919 & 0.6408 & 0.6137 & \bf\textcolor{red}{0.8642} & \bf\textcolor{red}{0.8684} & \bf\textcolor{red}{0.6709} & \bf\textcolor{red}{11.8738}\\
        & &S& MUSIQ (CVPR, 2021) \cite{ke2021musiq} & 0.8147 & 0.8216 & 0.6113 & 0.6428 & \bf\textcolor{blue}{0.8513} & \bf\textcolor{blue}{0.8422} & \bf\textcolor{blue}{0.6517} & \bf\textcolor{blue}{11.9566}\\
        & &T& StairIQA (JSTSP, 2023) \cite{sun2023blind}& \bf\textcolor{blue}{0.8527} & \bf\textcolor{blue}{0.8549} & \bf\textcolor{blue}{0.6563} & \bf\textcolor{blue}{0.5421} & 0.8052 & 0.8271 & 0.6123 & 11.8403\\
        & &U& \textbf{DHQI+SVR (Proposed)} & \bf\textcolor{red}{0.8559} & \bf\textcolor{red}{0.8644} & \bf\textcolor{red}{0.6747} & \bf\textcolor{red}{0.5104} & \bf{0.8313} &  \bf{0.8459} & \bf{0.6443} & \bf{11.2526}  \\ 
        \hline
    \end{tabular}
    \label{tab:performance}
    \vspace{-0.4cm}
\end{table*}

\subsection{Validation Setup}
\subsubsection{Benchmark Databases}
In addition to the proposed SJTU-H3D database, we have incorporated the digital human quality assessment (DHHQA) database \cite{zhang2023perceptual} as an additional resource for benchmark validation. The DHHQA database comprises a total of 55 scanned digital human heads that serve as reference samples, along with 1,540 labeled distorted digital human heads. These distorted samples have been intentionally degraded through the introduction of noise and compression/simplification.

\subsubsection{$k$-fold Cross-Validation}
To ensure robust evaluation, we adopt a $k$-fold cross-validation strategy. This approach involves dividing the database into $k$ equally sized folds. The model is then trained on $k$-1 of these folds and subsequently tested on the remaining fold. This process is repeated $k$ times, with each fold being used as the test set once. By averaging the performance across these $k$ iterations, we obtain a more reliable estimate of the model's effectiveness, minimizing the impact of random variations.
For both the SJTU-H3D and DHHQA databases, we have selected a value of $k=5$ to conduct the $k$-fold cross-validation, ensuring a balanced evaluation across multiple subsets. It's worth mentioning that there is no content overlap between the training and testing folds.

To facilitate a direct and fair comparison between zero-shot and supervised methods, we validate their performance in the following way. Zero-shot methods are directly applied to the testing folds, as they do not require any training. The performance is then averaged across the testing folds and reported as the final performance. On the other hand, supervised methods undergo training on the training folds and are subsequently tested on the testing folds. Similar to zero-shot methods, the average performance is calculated and reported as the final performance. Adopting this methodology enables a direct and unbiased comparison of the performance between zero-shot and supervised methods, providing insights into their respective strengths and limitations.

\subsubsection{Implemetation Details}
The cube-like projection process described in Section \ref{sec:pre} is conducted with the assistance of \textit{open3d} \cite{Zhou2018} library with a resolution of 1080P. The white background is cropped out. The projections are downsampled to 224$\times$224 as the input of the CLIP \cite{radford2021learning} image encoder. The ViT-B-32 \cite{dosovitskiy2020image} backbone with LAION-2B \cite{schuhmann2022laion} pretrained weights is utilized as the CLIP model. To fit the DHHQA database, we replace the suffix ``\textit{projection of 3d human model}" as described in Equation \ref{equ:text} with ``\textit{projection of 3d human face}". The scale parameter $c_1$ constant described in Section \ref{sec:niqe} is set as 100. The supervised training of the proposed DHQA method is conducted with the Support Vector Regression (SVR) model with RBF kernel.

The official source code is used for the competitors and default parameters are maintained. The default $5$-fold cross-validation is strictly followed for the competitors to make the comparison fair. In addition, the predicted scores of all the methods are followed by a five-parameter logistic regression to map the scores to the MOS scale.

\subsubsection{Evaluation Crieria}
Four mainstream criteria are employed for evaluation, which include Spearman Rank Correlation Coefficient (SRCC), Pearson Linear Correlation Coefficient (PLCC), Kendall’s Rank Order Correlation Coefficient (KRCC), and Root Mean Squared Error (RMSE). SRCC gauges the correlation of ranks, PLCC represents linear correlation, KRCC reflects the likeness of the orderings, while RMSE measures the quality prediction accuracy. A top-performing model should have SRCC, PLCC, and KRCC values that approach 1 and RMSE values close to 0.

\subsection{Competitors Selection}
The competitors' selection is conducted to ensure high diversity, which includes the zero-shot FR  methods, zero-shot NR methods, and the supervised NR methods.
\subsubsection{Zero-shot FR Methods}
We consider several classical projection-based FR methods: PSNR, SSIM \cite{ssim}, MS-SSIM \cite{wang2003multiscale}, and GMSD \cite{xue2013gradient}. These methods are applied to the six perpendicular projections, and the resulting scores are averaged and recorded. Additionally, we incorporate three popular point-based FR metrics proposed by MPEG: PSNR$_{p2po}$ \cite{tian2017evaluation}, PSNR$_{p2pl}$ \cite{tian2017updates}, and PSNR$_{yuv}$ \cite{mekuria2016design}. For the purpose of validation, we convert the digital human models into point clouds. Furthermore, we utilize G-LPIPS* \cite{nehme2022textured}, which is a projection-based FR metric modified from LPIPS \cite{zhang2018unreasonable} and is designed for textured meshes. The official pretrained weights are employed for this metric.
\subsubsection{Zero-shot NR Methods}
These methods comprise CPBD \cite{narvekar2011no}, pretrained BRISQUE* \cite{mittal2012brisque}, NIQE \cite{mittal2012making}, and IL-NIQE \cite{zhang2015feature}.
\subsubsection{Supervised NR Methods}
These methods encompass handcrafted approaches such as BRISQUE \cite{mittal2012brisque}, NFERM \cite{gu2014using}, and BMPRI \cite{min2018blind}, which are supervised using the Support Vector Regression (SVR) model. Additionally, we include deep learning-based methods, namely DBCNN \cite{zhang2018blind}, HyperIQA \cite{su2020blindly}, MUSIQ \cite{ke2021musiq}, and StaiIQA \cite{sun2023blind}, which have been retrained for our evaluation.

\begin{table*}[!htp]
    \centering
    \setlength{\tabcolsep}{4.8pt}
    \caption{SRCC \& PLCC performance comparison with zero-shot methods for different types of distortions on the SJTU-H3D database. }
    \begin{tabular}{l:l|ccccccc|ccccccc}
    \toprule
        \multirow{2}{*}{Ref} & \multirow{2}{*}{Method} & \multicolumn{7}{c|}{SRCC$\uparrow$} & \multicolumn{7}{c}{PLCC$\uparrow$} \\ \cline{3-16}
        &&GN & CN & FS & PC & UMC & TD & TC &GN & CN & FS & PC & UMC & TD & TC \\ \hline
        \multirow{4}{*}{FR} & PSNR &0.5728 & 0.4929 & 0.5853 & 0.7815 & 0.7871 & 0.2882 & 0.2901 & 0.5691 & 0.4770 & {0.5610} & 0.7811 & 0.7720 & 0.2702 & 0.1881 \\
        & SSIM & 0.6557 & \bf\textcolor{blue}{0.8218} & \bf\textcolor{red}{0.6961} & 0.7300 & 0.6224 & 0.6453 & 0.6222 &  0.6724 & \bf\textcolor{blue}{0.8024} & \bf\textcolor{red}{0.6621} & 0.7455 & 0.6035 & 0.7015 & 0.6074\\
        & MS-SSIM & 0.3494 & 0.1390 & 0.1280 & 0.3942 & 0.1988 & 0.0020 & 0.0719 &0.3378 & 0.0667 & 0.0766 & 0.3626 & 0.1883 & 0.0413 & 0.0364 \\
        & GMSD & 0.3369 & 0.1955 & 0.0593 & 0.6352 & 0.5387 & 0.0989 & 0.1680 &0.3245 & 0.2419 & 0.0962 & 0.6225 & 0.4749 & 0.1021 & 0.0931 \\ 
        &PSNR$_{p2po}$ & \bf\textcolor{red}{0.8771} & / & {0.6177} & 0.8112 & 0.4136 & / & / & \bf\textcolor{red}{0.8772} & / & 0.5562 & 0.7703 & 0.5853 & / & / \\
        &PSNR$_{p2pl}$ & \bf\textcolor{blue}{0.8365} & / & 0.2444 & 0.8180 & 0.4248 & / & / &\bf\textcolor{blue}{0.7937} & / & 0.3126 & 0.6823 & 0.5780 & / & / \\
        &PSNR$_{yuv}$ & 0.5670 & \bf\textcolor{red}{0.8390} & 0.3163 & \bf\textcolor{blue}{0.8465} & \bf\textcolor{blue}{0.8547} & \bf\textcolor{blue}{0.7211} & 0.8039 &0.6419 & \bf\textcolor{red}{0.8472} & 0.2329 & \bf\textcolor{blue}{0.8288} & \bf\textcolor{blue}{0.8354} & 0.7585 & 0.7996 \\
        &G-LPIPS* &0.7633 & 0.7258 & \bf\textcolor{blue}{0.6307} & 0.8097 & 0.7268 & 0.7199 & 0.7491 &0.7421 & 0.7106 & \bf\textcolor{blue}{0.6228} & 0.7787 & 0.7093 & 0.7370 & 0.7380\\ \hdashline
        \multirow{4}{*}{NR} & CPBD & 0.0646 & 0.2615 & 0.1366 & 0.3629 & 0.7891 & 0.5614 & 0.5135 &0.0323 & 0.1786 & 0.0619 & 0.3945 & 0.7765 & 0.6376 & 0.5178 \\
        & BRISQUE* & 0.1597 & 0.3809 & 0.0004 & 0.3151 & 0.6208 & 0.2989 & 0.4957 &0.1532 & 0.3930 & 0.0135 & 0.3240 & 0.6220 & 0.3697 & 0.4968\\
        & NIQE & 0.0266 & 0.6455 & 0.0822 & 0.2535 & {0.8002} & 0.6682 & 0.8233 &0.0275 & 0.5946 & 0.0648 & 0.3950 & 0.7499 & \bf\textcolor{blue}{0.7862} & 0.8311 \\
        & IL-NIQE &0.2667 & 0.7578 & 0.1286 & 0.4513 & 0.5666 & 0.6130 & \bf\textcolor{red}{0.8582} &0.2346 & 0.7099 & 0.1457 & 0.4416 & 0.6311 & 0.7547 & \bf\textcolor{red}{0.8463} \\
        & \bf{DHQI} & \bf{0.8075} & \bf{0.7880} & \bf{0.4715} & \bf\textcolor{red}{0.8590} & \bf\textcolor{red}{0.8676} & \bf\textcolor{red}{0.7798} & \bf\textcolor{blue}{0.8395} &\bf{0.7934} & \bf{0.7681} & \bf{0.4686} & \bf\textcolor{red}{0.8926} & \bf\textcolor{red}{0.8677} & \bf\textcolor{red}{0.8407} & \bf\textcolor{blue}{0.8322}
        \\
    \bottomrule
    \end{tabular}
    \label{tab:detail}
    \vspace{-0.4cm}
\end{table*}

\begin{table}[!htp]
    \centering
    \caption{Ablation study results on the SJTU-H3D and DHHQA databases. }
    \begin{tabular}{c:c:c|cc|cc}
    \toprule
    \multicolumn{3}{c|}{Quality Measure} & \multicolumn{2}{c|}{SJTU-H3D} & \multicolumn{2}{c}{DHHQA} \\ \hline
      $Q_A$  & $Q_N$ & $Q_G$ & SRCC$\uparrow$ & PLCC$\uparrow$ & SRCC$\uparrow$ & PLCC$\uparrow$ \\ \hline
      \checkmark & $\times$ & $\times$  & 0.5631 &  0.6056 & 0.6379 &  0.6639\\
      $\times$ & \checkmark & $\times$  & 0.4404 &  0.5575 & 0.4182 &  0.4581\\
      $\times$ & $\times$ & \checkmark   & 0.1771 &  0.3782 & 0.5446 &  0.7178\\ \hdashline
      $\times$ & \checkmark & \checkmark & 0.6817 &  0.6990  & 0.6244 &  0.7174\\
      \checkmark & $\times$ & \checkmark  & 0.6353 &  0.6432  & \bf\textcolor{blue}{0.7304} &  \bf\textcolor{blue}{0.7651}\\
      \checkmark & \checkmark & $\times$   & \bf\textcolor{blue}{0.7435} &  \bf\textcolor{blue}{0.7480} & 0.7245 &  0.7349\\  \hdashline
      \checkmark & \checkmark & \checkmark & \bf\textcolor{red}{0.8376} &  \bf\textcolor{red}{0.8276} & \bf\textcolor{red}{0.7882} &  \bf\textcolor{red}{0.8070}\\
    \bottomrule
    \end{tabular}
    \label{tab:abl}
    \vspace{-0.4cm}
\end{table}

\subsection{Performance Discussion}
The overall performance on the SJTU-H3D and DHHQA databases are exhibited in Table \ref{tab:performance}, from which we can draw several conclusions. 

\subsubsection{Zero-shot Performance} 
a) Among all the zero-shot methods compared on the SJTU-H3D database, the DHQI method demonstrates superior performance and outperforms them all. Additionally, it proves to be competitive even when compared to FR  metrics on the DHHQA database.
b) Nevertheless, the FR metrics that exhibit the highest performance on the DHHQA database, namely MS-SSIM \& GMSD, suffer significant performance degradation when applied to the SJTU-H3D database. This decline suggests that these metrics lack robustness in handling diverse digital human content.
c) In contrast, all the competing zero-shot NR methods consistently exhibit lower performance compared to the proposed DHQI method. The reason for this disparity lies in the focus of these methods on addressing low-level distortions, which restricts their ability to effectively capture and model high-level semantic quality representations. By leveraging the semantic affinity quality measure, the DHQI method can enhance the performance of zero-shot NR approaches even further.

\subsubsection{Supervised Performance}
Due to the significant advancements achieved by deep neural networks, deep learning-based methods such as HyperIQA and MUSIQ have demonstrated superior performance compared to traditional handcrafted methods. Despite this, the proposed DHQI method, which is solely supervised by Support Vector Regression (SVR) model, achieves the top-ranking performance on the SJTU-H3D database.
One notable advantage of the proposed supervised DHQI index is its cost-effectiveness in terms of time and computational resources. The calibration process of an SVR model requires considerably less time and computational overhead compared to training and optimizing deep neural networks. This attribute enhances the practical viability and efficiency of the proposed DHQI method.

\subsection{Distortion-specific Performance}
To investigate the specific effects of zero-shot methods, we present the distortion-specific performance in Table \ref{tab:detail}, from which we can several observations:
a) The proposed DHQI method achieves first place in three types of distortions: PC, UMC, and TD, which demonstrates its effectiveness in handling these distortions.
b) The point-based methods proposed by MPEG exhibit high sensitivity to noise-related distortions. This can be attributed to the direct impact of geometry and color noise on the point-level quality characteristics. Additionally, the PSNR$_{yuv}$ metric demonstrates a strong discriminative ability in distinguishing quality differences within CN, PC, UMC, and TD distortions. However, it is less effective in handling cross-distortion content from a general perspective (its overall SRCC performance is just 0.5247).
c) The zero-shot NR methods NIQE and IL-NIQE show competitive performance for UMC, TD, and TC distortions. This can be attributed to the fact that UMC and TD distortions introduce blurring effects to digital human projections, which aligns with the strengths of these methods. TC distortion, on the other hand, introduces typical JPEG artifacts to digital humans, which can be easily quantified by these methods as well.
d) FS distortion proves to be the most challenging distortion to evaluate. This is due to the fact that the MOS distribution for FS distortion tends to be more centered, as shown in Fig. \ref{fig:mos}, indicating a more fine-grained quality level that is less distinctive. FS distortion primarily causes digital humans to exhibit more geometric characteristics, which may lead to small differences in NSS reflected by the projections and result in the poor performance of NIQE and IL-NIQE.
Despite the less competitive performance of the proposed method in handling FS distortion, it significantly advances the performance of NR methods in general.

\subsection{Ablation Study}
In this section, we present an analysis of the effects of different quality measures: $Q_A$, $Q_N$, and $Q_G$, on the experimental performance. The combinations of these quality measures are tested, and the results are summarized in Table \ref{tab:abl}. Throughout the experiments, we maintain the default experimental setup.
Table \ref{tab:abl} clearly demonstrates that among the single quality measures, $Q_A$ achieves the highest performance. This finding indicates a strong correlation between quality-aware semantic affinity and the visual quality of digital humans. It suggests that considering the quality of semantic representations is crucial for accurately assessing the visual fidelity of digital human models.
Furthermore, excluding any of the three quality measures leads to a drop in performance compared to utilizing all quality measures together. This observation implies that each quality measure contributes significantly to the final results. The effectiveness of the proposed framework is thereby validated by the consistent performance improvements achieved when all quality measures are incorporated.

%\subsection{Efficiency Discussion}
\begin{figure}
    \centering
    \subfigure[Zero-shot SJTU-H3D]{\centering \includegraphics[width = 0.45\linewidth]{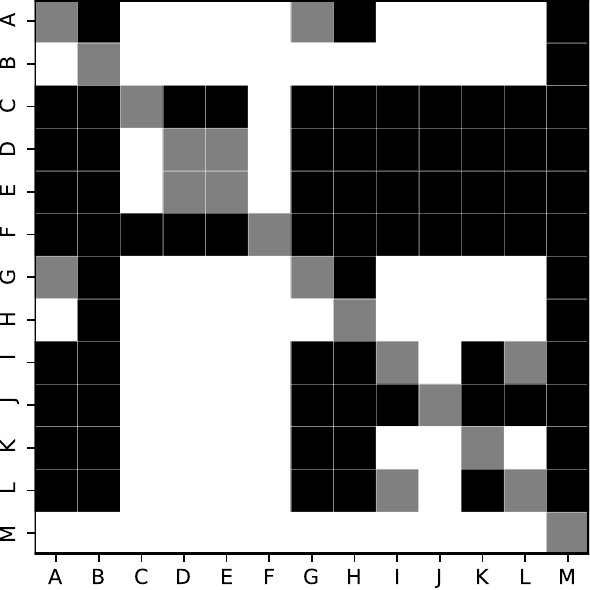}}
    \subfigure[Zero-shot DHHQA]{\centering \includegraphics[width = 0.45\linewidth]{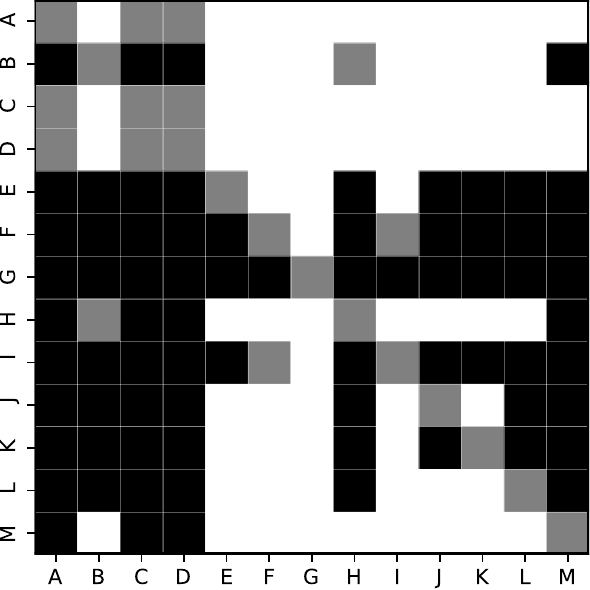}}
    \subfigure[Supervised SJTU-H3D]{\centering \includegraphics[width = 0.45\linewidth]{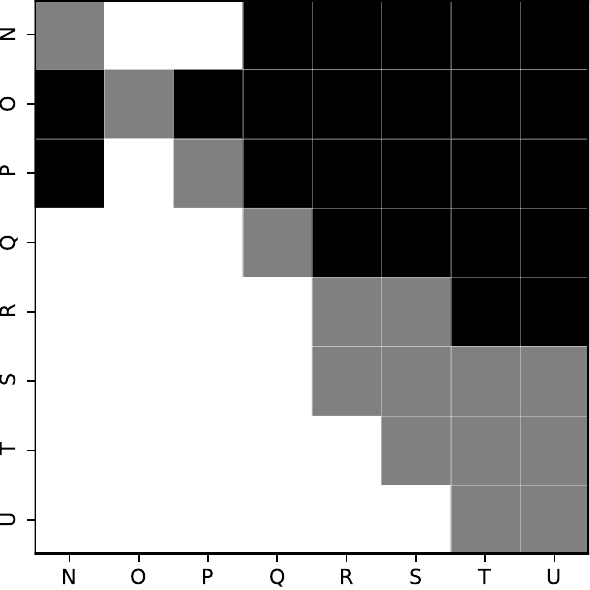}}
    \subfigure[Supervised DHHQA]{\centering \includegraphics[width = 0.45\linewidth]{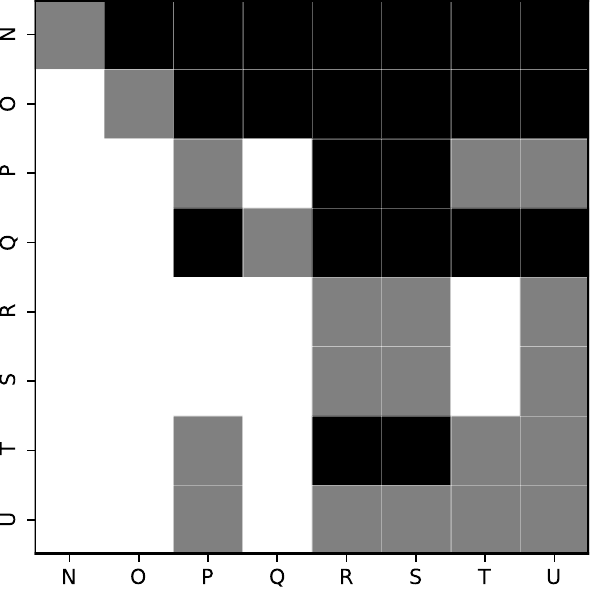}}
    \caption{Statistical test results of the proposed method and compared methods on the SJTU-H3D and DHHQA databases. A black/white block means the row method is statistically worse/better than the column one. A gray block means the row method and the column method are statistically indistinguishable. The methods are denoted by the same index as in Table \ref{tab:performance}.}
    \label{fig:heatmap}
    \vspace{-0.4cm}
\end{figure}

\subsection{Statistical Test}
To further analyze the performance of the proposed method, we conduct the statistical test in this section. We follow the same experiment setup as in \cite{statistic-test} and compare the difference between the predicted quality scores with the subjective ratings. All possible pairs of models are tested and the results are listed in Fig. \ref{fig:heatmap}. 
Our method demonstrates remarkable superiority over 12 zero-shot methods and 5 supervised methods when compared on the SJTU-H3D database. On the DHHQA database, our method exhibits substantial outperformance compared to 9 zero-shot methods and 3 supervised methods.

\section{Conclusion}
The increasing applications of digital humans across various domains have highlighted the need for comprehensive quality assessment studies. However, the limited availability of comprehensive digital human quality assessment (DHQA) databases has posed challenges in this area. To address this gap, we have introduced the SJTU-H3D subjective quality assessment database, specifically designed for full-body digital humans. This database consists of 40 high-quality reference digital humans and 1,120 labeled distorted counterparts created with seven types of distortions. 
Nonetheless, the scarcity of suitable DHQA databases remains a hindrance to the development of data-driven methods. To overcome this limitation and enhance generalization capabilities, we propose a zero-shot DHQA approach that focuses on no-reference (NR) scenarios. Our approach leverages semantic and distortion features obtained from projections, as well as geometry features derived from the mesh structure of digital humans. 
The proposed DHQI not only serves as a robust baseline for DHQA tasks but also facilitates advancements in the field. We hope our work can contribute to the establishment of effective evaluation frameworks and methodologies for digital humans, enabling their widespread application in diverse domains.

\bibliographystyle{IEEEtran}
\bibliography{journal}
\end{document}